# Highly Distorted Lattices in Chemically Complex Alloys Produce Ultra-Elastic Materials with Extraordinary Elinvar Effects

Q.F. He[1†], J.G. Wang[1,2†], H.A. Chen[3†], Z.Y. Ding[1], Z.Q. Zhou[1], L.H. Xiong[4], J.H. Luan[5], J.M. Pelletier[6], J.C. Qiao[1,6,7], Q. Wang[8], L.L. Fan[9], Y. Ren[10], Q.S. Zeng[4,11], C.T. Liu[1,5], C.W. Pao[3*], D.J. Srolovitz[1,5*], Y. Yang[1,5*]

[1]Department of Mechanical Engineering, City University of Hong Kong, Kowloon Tong, Kowloon, Hong Kong, China

[2]College of Mechanical Engineering, Dongguan University of Technology, Dongguan, Guangdong, China

[3]Research Center for Applied Sciences, Academia Sinica, Taipei 11529, Taiwan

[4]Jiangsu Key Laboratory of Advanced Metallic Materials, School of Materials Science and Engineering, Southeast University, Nanjing 211189, China

[5]Department of Materials Science and Engineering, City University of Hong Kong, Kowloon Tong, Kowloon, Hong Kong, China

[6]Université de Lyon, MATEIS, UMR CNRS5510, Bat. B. Pascal, INSA-Lyon, F-69621 Villeurbanne Cedex, France

[7]School of Mechanics, Civil Engineering and Architecture, Northwestern Polytechnical University, Xi'an 710072, China

[8]Laboratory for Structures, Institute of Materials, Shanghai University, Shanghai 200072, China

[9]College of Physics and Materials Science, Tianjin Normal University, Tianjin 300387, China

[10]X-ray Science Division, Argonne National Laboratory, Lemont, IL 60439, USA

[11]Center for High Pressure Science and Technology Advanced Research, Pudong, Shanghai 201203, China

*Correspondence to: yonyang@cityu.edu.hk (YY); cwpao@gate.sinica.edu.tw (CWP) and srol@cityu.edu.hk (DS)

†Authors contribute equally to this work.

**Abstract:**

Conventional crystalline alloys usually possess a low atomic size difference in order to stabilize its crystalline structure. However, in this article, we report a single phase chemically complex alloy which possesses a large atomic size misfit usually unaffordable to conventional alloys. Consequently, this alloy develops a rather complex atomic-scale chemical order and a highly distorted crystalline structure. As a result, this crystalline alloy displays an unusually high elastic strain limit (~2%), about ten times of that of conventional alloys, and an extremely low internal friction ($< 2 \times 10^{-4}$) at room temperature. More interestingly, this alloy firmly maintains its elastic modulus even when the testing temperature rises from room temperature to 900 K, which is unmatched by the existing alloys hitherto reported. From an application viewpoint, our discovery may open up new opportunities to design high precision devices usable even under an extreme environment.

**Main Text:**

The development of high performance ultra-elastic metals with superb strength, a large elastic strain limit and temperature insensitive elastic modulus (Elinvar effect) is important in a variety of industrial applications, from actuators to medical devices to high precision instruments (*1, 2*). There are no generally accepted design principles for such ultra-elastic alloys. Plastic deformation through dislocations, twinning and/or grain boundary sliding commonly limit the elastic strain of bulk crystalline metals to ~0.2% at room temperature (*3*). Because bulk amorphous alloys lack crystal defects, they may achieve ~2% elastic strain at room temperature (*4*). However, the high cooling rate required to form most amorphous metallic alloys provides severe limits on the size that can be produced, thereby seriously hindering their industrial application (*5*). On the other hand, shape memory alloys may achieve elastic strain limit up to several percent (*3*). However, this high strain limit is associated with reversible martensitic transformations which are accompanied by large mechanical hysteresis and energy dissipation (*3*); such deformation is pseudoelastic rather than elastic. Apart from bulk amorphous and shape memory alloys, we notice that gum metals can also achieve a large elastic strain limit (~ 2%) (6-8). It was hypothesized that the super-elasticity in gum metals is associated with a dislocation and transformation-free deformation mechanism (6-8). However, recent experiments clearly revealed the large elastic strain limit in these materials is associated with martensitic transformations and dislocation activity prior to overall yielding (9-11) and that this leads to significant mechanical hysteresis/energy dissipation (12, 13). In this work, we report our development of a new class of alloys that exhibit linear elastic response to very high strains without hysteresis and elastic properties which are temperature-insensitive (to high temperature). This mechanical property profile is ideal for a wide-range of applications, yet has never been reported before (14, 15).

Concentrated multi-component alloys (i.e., high entropy alloys, HEAs) (*16*) have recently provided a new route to design new metallic alloys with superior mechanical and physical properties (*17-19*). Early HEA research focused mainly on the formation of single-phase solid solutions by minimizing the overall atomic size misfit in these alloys. A widespread view is that large atomic size misfit favors amorphous phase formation (*20, 21*). Here, we report the development of a chemically complex, single phase alloy in which the atomic size misfit is too large according to most solid solution alloy design rules. This misfit introduces extremely large atomic-scale distortions within a single crystalline phase. The resultant alloy exhibits an ultra-large elastic strain limit (~2%) and negligible internal friction at room temperature. Even more surprising is that the elastic properties of this chemically complex alloy are nearly constant even up to 900 K. This near temperature-independence of the elastic moduli is known as the Elinvar effect; this effect is stronger than that in any other metallic alloys reported to-date (6).

We prepared the chemical complex alloy $Co_{25}Ni_{25}(HfTiZr)_{50}$ (atomic %) via high rate directional solidification method. Unlike conventional single phase HEAs (20), the atomic size difference in $Co_{25}Ni_{25}(HfTiZr)_{50}$ alloy is extremely large, ~11%, relative to standard approaches used for single phase alloy design (21) (see the **Supplementary Text**). The established alloy phase rules/theories (20, 22) suggest that such large atomic size misfit will destabilize the crystalline structure, leading to either phase separation or the formation of a single phase, amorphous structure (20). X-ray diffraction (XRD) and electron backscattered diffraction (EBSD) (Fig. 1A) demonstrate the $Co_{25}Ni_{25}(HfTiZr)_{50}$ alloy is a single-phase crystal with a nominally body-centered cubic (BCC) like structure. A low magnification elemental mapping and TEM imaging show that the $Co_{25}Ni_{25}(HfTiZr)_{50}$ alloy is a single phase ordered BCC-like structure without any phase separation (**Fig. S1**). A more detailed 3D atom probe tomography (APT) characterization demonstrates that this $Co_{24.9}Ni_{24.9}Hf_{17.4}Ti_{17.3}Zr_{15.5}$ alloy is chemically homogeneous (spatial resolution ~ 1 nm) with a nearly random elemental distribution (see Fig. 1B); the constituent element distribution is consistent with the binomial distribution characteristic of a typical random alloy, as discussed in Ref. (23) (Fig. 1C).

Aberration corrected high-angle annular dark field scanning transmission electron microscopy (HADDF-STEM) observations were employed to determine the detailed atomic structure along different crystal zone axes; [111], [011] and [001] (see Figs. 1D-F). Interestingly, sub-nanometer spatial resolution elemental energy-dispersive X-ray spectroscopy (EDS) reveals small, atomic-scale, chemical order. STEM-EDS images of single crystal $Co_{25}Ni_{25}(HfTiZr)_{50}$ along the [011] crystal zone axis (Fig. 1G) shows that Co and Ni have tend to occupy one sub-lattice while Hf and Ti another (Zr atoms are randomly distributed between these sub-lattices). These results suggest that our alloy, while nominally a random BCC like structure, exhibits some B2 ordering.

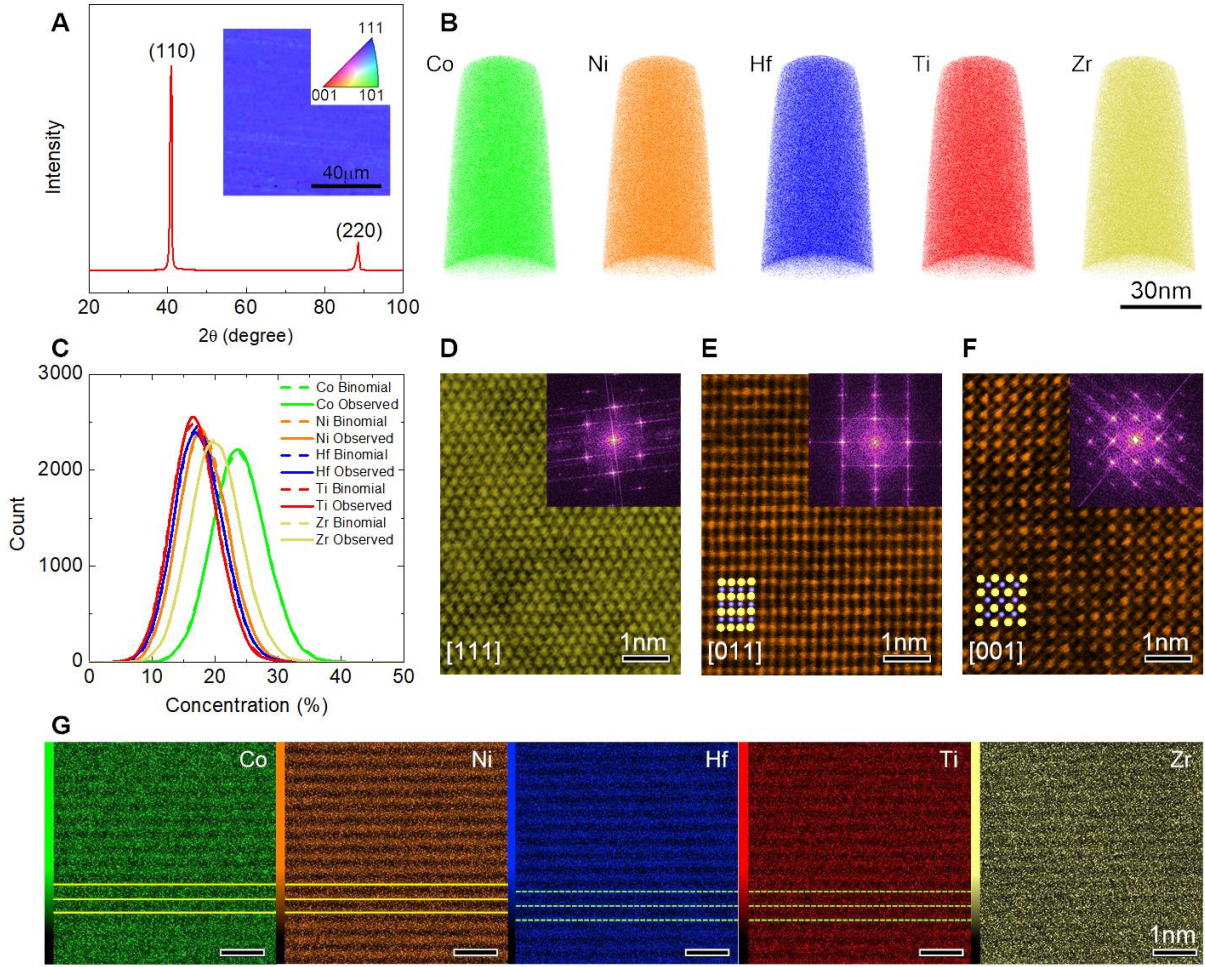

**Fig. 1. Microstructure of single crystal Co₂₅Ni₂₅(HfTiZr)₅₀ alloy.** (**A**) The XRD pattern and EBSD image of the single crystal Co₂₅Ni₂₅(HfTiZr)₅₀. Note that the EBSD image was taken from a single crystal compression sample while the XRD pattern corresponds to the strongly diffracted orientation of the single crystal. (**B**) The APT reconstructions of three-dimensional elementary distributions. (**C**) The statistical binomial frequency distribution of the constituent elements. All the experimental observed distribution curves match very well with the totally random distribution

curves. (**D-F**) The STEM-HAADF images from different crystal axes of [111], [011] and [001], respectively (insets show the fast Fourier transform FFT of these data patterns). (**G**) Detailed STEM-HADDF elemental maps show that some chemical ordering exists on the sub-nanometer scale.

We performed extensive density functional theory (DFT) calculations on the $Co_{25}Ni_{25}(HfTiZr)_{50}$ alloy (see **Methods**) to provide further insight into this structure. We constructed three structurally distinct models ( illustrated in Fig. S2A): (I) Disordered: elements assigned at random to all BCC lattice sites, (II) Ordered: sites in sublattice A of the B2 structure (corner sites of BCC unit cell) occupied at random with {Co, Ni} and sites in sublattice B (center sites of the BCC unit cell) are occupied at random with {Hf, Ti, Zr}, and (III) Partially Ordered: as per Model III but with 25% of Zr atoms from sublattice A exchanging with Co and Ni atoms on sublattice B – as suggested by our STEM observations. Atomic relaxation (T=0 K) of these models demonstrate that Model I is unstable (i.e., spontaneously forming an amorphous structure), in agreement with the established phase stability rules (*20*). In contrast, Models II and Model III remain stable throughout the relaxation. This indicates that the large atomic size misfit in the $Co_{25}Ni_{25}(HfTiZr)_{50}$ alloy can be accommodated by atomic-scale chemical ordering.

We constructed up to 150 realizations of each of the Models via an atom swapping algorithm followed by relaxation and calculated the corresponding potential energy. On average, Models I and III have a higher potential energy than Model II (Fig. S2B). This suggests that the atomic structure (i.e., Model III) of the $Co_{25}Ni_{25}(HfTiZr)_{50}$ alloy obtained by a high rate directional solidification is metastable with respect to a more fully ordered structure (Model II). Similar results were also obtained at room temperature 300K (Fig. S3). The metastable amorphous structure (Model I) would be also experimentally obtained via magnetron sputtering (Fig. S4). As a result of the large atomic size difference, the lattice structure of $Co_{25}Ni_{25}(HfTiZr)_{50}$ is significantly distorted. To quantify the lattice distortion, we calculated the lattice distortion-induced volumetric and von Mises shear strains in Model II and III (see **Methods**). As seen in Fig. 2A, the volumetric and von Mises shear strains calculated for Model III is widely distributed. Note that the average von Mises (shear) strain obtained from Model III is very large (~9%); much larger than in conventional alloys and many HEAs (e.g., Co-Cr-Fe-Mn-Ni alloy) (*24*). We obtained similar results for Model II, in which the average von Mises strain is 3-7% (Fig. S5). The local structural environment around each atom was determined using the polyhedral template matching (PTM) method(*25*). Figure 2B shows the distributions of root mean square deviation (RMSD) of the central atom and its neighbors in first coordination shell compared with the counterparts in the reference lattice in Model III. Evidently, the majority of the atomic clusters can be characterized as distorted local BCC packing (RMSD=0.1) while a noticeable portion of the atomic clusters is akin to local hexagonal close packing (HCP) or even simple cubic (SC) packing as a result of remarkable lattice distortion. These calculations deliver a strong message that, in addition to the rather complex atomic-scale chemical ordering, the $Co_{25}Ni_{25}(HfTiZr)_{50}$ alloy also possesses a highly distorted lattice. Such a unique combination of structural features has never been reported for conventional alloys or other HEAs.

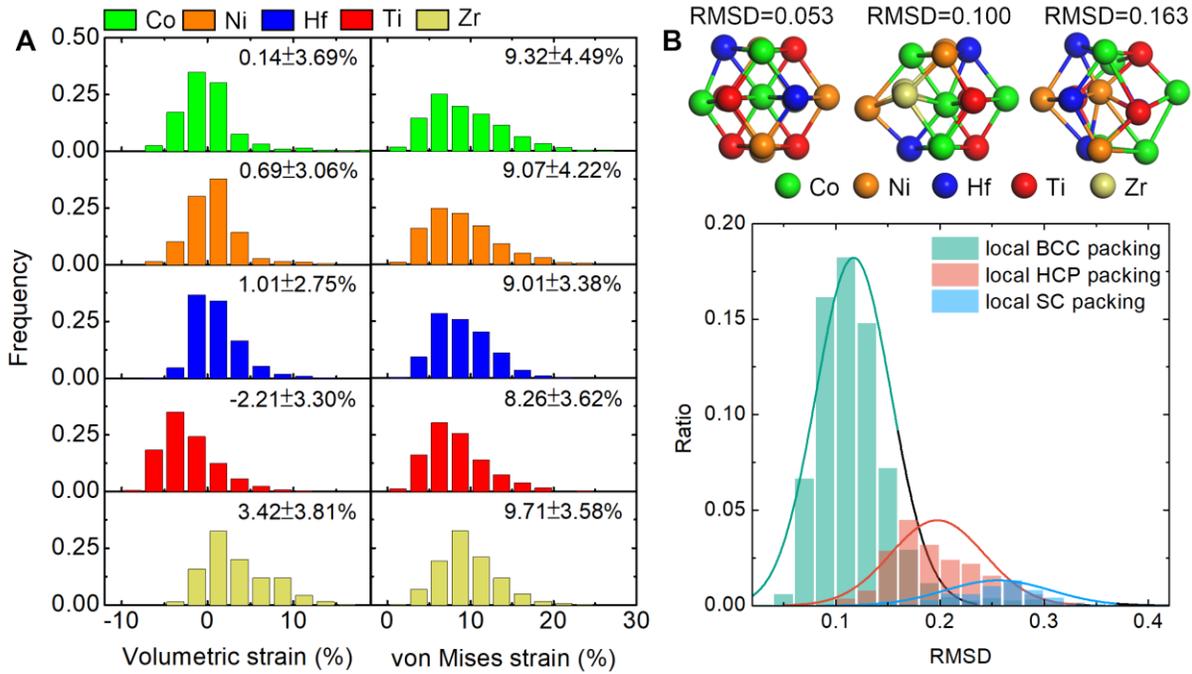

**Fig. 2. DFT calculation revealed local atomic structures of Co$_{25}$Ni$_{25}$(HfTiZr)$_{50}$ alloy.** (**A**) The atomic volumetric and von Mises shear strain distribution of Model III structures which is in line with the STEM observations. (**B**) The RMSD distributions of atomic clusters in Model III structures. The top parts are several representative BCC atomic clusters with different RMSD values. The RMSD is a measure of the spatial deviation from the perfect structure template.

A series of compression tests at different scales were conducted to characterize the mechanical properties of the Co$_{25}$Ni$_{25}$(HfTiZr)$_{50}$ alloy at room temperature. Figure 3A shows the compressive stress-strain curve of the single crystal along the [111] direction and polycrystalline (obtained by arc melting - see **Methods**) Co$_{25}$Ni$_{25}$(HfTiZr)$_{50}$. As seen in Fig. S6, the polycrystalline Co$_{25}$Ni$_{25}$(HfTiZr)$_{50}$ alloy possesses a similar atomic structure to its single crystal counterpart. Interestingly, regardless of their microstructural difference, both the single crystal and polycrystalline alloy exhibited almost the same yield strength (1.92 GPa for single crystal and 1.96

GPa for polycrystal specimen see Fig. S7). The Young's modulus (extracted from a linear fit to the stress-strain curves at small strains) is ~99GPa for the single crystal sample and ~94GPa for the polycrystal sample (see Fig. S7). The $Co_{25}Ni_{25}(HfTiZr)_{50}$ alloy exhibited a very high elastic strain limit (~2%) which is determined from the relation of $\varepsilon_E = \sigma_y / E$; this is nearly an order of magnitude larger than that of conventional crystalline metals (3). We verified that the microstructure and mechanical properties of the $Co_{25}Ni_{25}(HfTiZr)_{50}$ were insensitive to annealing at 1273K for different time durations (see Fig. S8). In addition, we fabricated 39 single crystal $Co_{25}Ni_{25}(HfTiZr)_{50}$ micropillars with diameters ranging from 0.3 to 5μm for micro-compression studies (see Fig. S9A). Typical stress strain curves (see the inset in Fig. 3A) shows that the micropillar yield strength is slightly larger than that of the bulk sample, which may be attributed to surface effects (Weibull statistics) (26) (Fig. S10). To extract Young's modulus values from these micropillar compression tests, we correct for pillar taper angle and substrate effects following the approach described in Ref. (27). The Young's modulus of $Co_{25}Ni_{25}(HfTiZr)_{50}$ alloy is obtained as 106±10GPa after averaging the data from 39 micropillars (Fig. S9B), which is consistent with the results of bulk tests.

Figure 3B shows the room-temperature yield strength versus pillar diameter obtained from the micropillars compression. To compare with the literature data in different systems (including single crystal face centered cubic (FCC)(28-31), single crystal BCC(32) and bulk metallic glasses, BMGs (33-35)), we normalize the yield strengths by their corresponding elastic moduli. For consistency with the literature data, we report the yield strength of the single crystal $Co_{25}Ni_{25}(HfTiZr)_{50}$ micropillars as the flow stress at the 5% nominal strain. We can characterize the dependence of the yield strength on pillar diameter as a power law with an exponent given by the slope *m* in a log-log plot (Fig. 3B and Fig. S9C). While this exponent is much larger for FCC ($m$ = -0.6) than BCC ($m$ = -0.35) single crystals, both are larger than that for BMGs ($m$ = -0.11). Remarkably, the yield strength of Co$_{25}$Ni$_{25}$(HfTiZr)$_{50}$ single crystal is nearly pillar size-independent ($m$ ~ -0.07). More importantly, however, the single crystal Co$_{25}$Ni$_{25}$(HfTiZr)$_{50}$ possesses the highest (normalized) strength or elastic strain limit amongst all of the crystalline materials and is comparable to that of the BMGs. These microcompression yield strength results are consistent with our nanoindentation data, shown in Fig. S11.

To understand the mechanisms underlying the observed ultra-high elastic strain limit, we performed dynamic mechanical spectroscopy analyses in an inverted torsion pendulum in a high vacuum atmosphere (see **Methods**). As shown in Fig. S12, the loss factor (tan$\delta$) for Co$_{25}$Ni$_{25}$(HfTiZr)$_{50}$ alloy is $2\times10^{-4}$ at room temperature. This low loss factor is smaller than those measured for a variety of BMGs (but similar to that of a Zr-Cu-Al BMG). While the BMG loss factors increase dramatically with increasing temperature (particularly near their glass transition temperature), the low Co$_{25}$Ni$_{25}$(HfTiZr)$_{50}$ alloy loss factor remains nearly constant over a very wide temperature range. Cyclic microcompression tests were conducted at several loading rates; unlike in the BMGs (*36*), no mechanical hysteresis was observed prior to yielding (Fig. S9D). In other words, the energy storage efficiency in the Co$_{25}$Ni$_{25}$(HfTiZr)$_{50}$ alloy is close to 100%, much higher than the hitherto reported ultra-elastic materials (*37*). Surprisingly, the elastic properties of this Co$_{25}$Ni$_{25}$(HfTiZr)$_{50}$ alloy exceed those of most conventional metallic materials, as seen in the tan$\delta$ versus elastic strain limit plot in Fig. 3C. This large elastic limit and low loss factor were directly visualized through a series of experiments in which a steel ball bounces against a Co$_{25}$Ni$_{25}$(HfTiZr)$_{50}$ alloy plate and plates of a Cu$_{50}$Zr$_{45}$Al$_5$ metallic glass (MG), a B2 NiAl alloy, and a commercial stainless steel (see **Supplementary Video** and Fig. S13); these observations demonstrate the lower mechanical loss and higher elastic limit compared with other structural metallic alloys.

Theoretically, both strong lattice distortion and atomic-scale chemical ordering may increase the lattice friction to dislocation motion (*38*). Nix et al. (*39, 40*) proposed that high lattice resistance to dislocation motion decreases the effect of sample size in single crystals. The near size-independence of the yield strength of our Co$_{25}$Ni$_{25}$(HfTiZr)$_{50}$ alloy may be attributed to a high lattice friction. However, given the chemical complexity of the Co$_{25}$Ni$_{25}$(HfTiZr)$_{50}$ alloy, we have been unable to directly determine a statistically meaningful lattice friction directly from first principles calculations. As an alternative, we made direct observations of the dislocation structures within a single crystal Co$_{25}$Ni$_{25}$(HfTiZr)$_{50}$ following 4% deformation and the preparation of TEM foils (see Fig. S14-15). Classical **g·b** contrast analysis (*41*) was employed to determine the burgers vector of the observed dislocations. While we expected to see <111> dislocation burgers vectors, characteristic of BCC materials, we observed that the dislocations in the Co$_{25}$Ni$_{25}$(HfTiZr)$_{50}$ alloy were predominantly of the <001> type (Fig. S14-15). While this is consistent with B2 crystal symmetry since the shortest translation vector in B2 is a$_0$<001>, (a$_0$/2)<111> dislocations are possible in B2-these would be separated by anti-phase boundaries, the energy of which increases with the strength of the ordering (*42*). We extracted the stress-dependence of the activation free energy for dislocation motion by measuring the activation volume (through stress relaxation experiments – see Fig. S16). The activation volume was measured to be 3.05**b**$^3$ , which is small for BCC metals and implies that deformation is controlled by the formation and movement of dislocation kink-pairs (*43*). These results suggest plastic yield in Co$_{25}$Ni$_{25}$(HfTiZr)$_{50}$ is associated with dislocation glide, rather than by other mechanisms such as nucleation. In other words, the high yield strength we observe in the Co$_{25}$Ni$_{25}$(HfTiZr)$_{50}$ alloy is associated with large dislocation glide barriers – presumably dominated by a strong lattice friction. The dominance of lattice friction is consistent with the very small size effect observed in the strength of the $Co_{25}Ni_{25}(HfTiZr)_{50}$ alloy (Fig. 3B).

A high lattice friction is consistent with a high Peierls stress $\tau_c$; i.e., the stress required to move a dislocation through the lattice. We estimate the Peierls stress of $Co_{25}Ni_{25}(HfTiZr)_{50}$ by extrapolating the high temperature yield strength data back to 0K. To this end, we performed a series of high temperature micropillar compression tests (see **Methods**). As seen in Fig. 3D, the yield strength of the micropillars declines with increasing temperature. By fitting the experimental data to the Peierls-Nabarro model $\sigma = \sigma_c \left[ 1 - \left( T/T_0 \right)^{2/3} \right]$, we find $\tau_c \simeq \sigma_c /2.5 = 2.4$ GPa (here $T$ is temperature and $T_0$ is a reference temperature). The large Peierls stress in $Co_{25}Ni_{25}(HfTiZr)_{50}$ is about forty times the average Peierls stress ($\tau_p \approx 54$ MPa) estimated by a simple rule of mixture from the elemental components in this alloy (see **Supplementary Text**). In a stochastic Peierls-Nabarro model (*38*), the high lattice friction stress may be attributed to local perturbations to the Peierls potential (caused by atomic scale randomness) and/or chemical ordering.

To examine if deformation induces any phase transformations before overall yielding (as in pseudo-elastic or shape memory alloys), we performed in-situ high energy X-ray diffraction (HEXRD) studies of single crystal $Co_{25}Ni_{25}(HfTiZr)_{50}$ during uniaxial [110] compressive loading and unloading. As shown in Fig. 4, the change in the d-spacing represents the loading (up to 2% strain) and unloading. Interestingly, no new phases appear during loading or unloading (Fig. 4A). This proves that the observed high elastic strain limit in $Co_{25}Ni_{25}(HfTiZr)_{50}$ is not associated with stress-induced phase transformations (like in pseudo-elastic and shape memory alloys), but rather arises from the high lattice friction, as proposed here. The cubic lattice parameter of the undeformed Co$_{25}$Ni$_{25}$(HfTiZr)$_{50}$ is 3.12Å. The linearity in the *d*-spacing versus stress, indicates that Co$_{25}$Ni$_{25}$(HfTiZr)$_{50}$ is linearly elastic up to 2% strain (Fig. 4B).

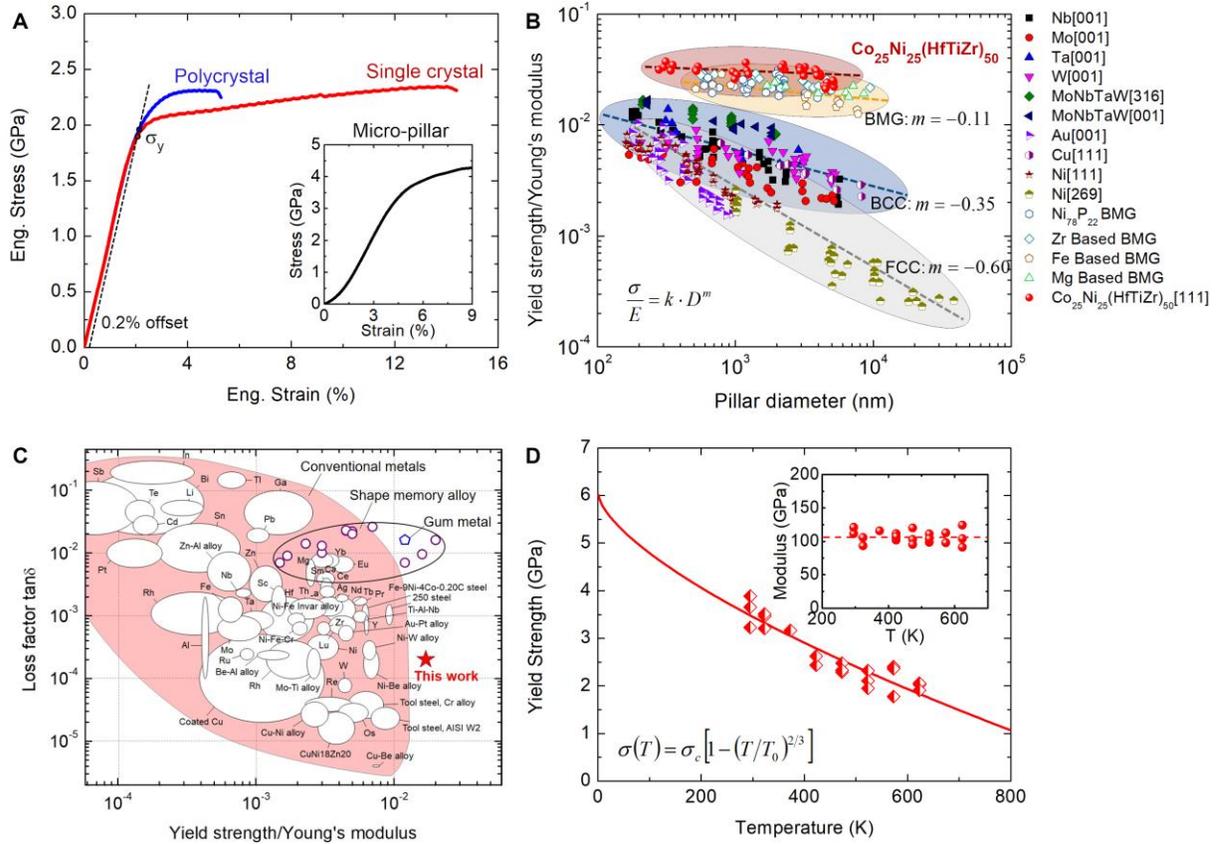

**Fig. 3. Mechanical properties of Co$_{25}$Ni$_{25}$(HfTiZr)$_{50}$ alloy.** (**A**) Compressive stress-strain curves for single crystal and polycrystalline Co$_{25}$Ni$_{25}$(HfTiZr)$_{50}$ alloys at room temperature. The inset shows the single crystal micropillar compressive stress-strain curve with top diameter about 1μm (from micropillar data set in **B**). The bulk single crystal compression was conducted along the [111] direction. (**B**) A comparison of the size-dependent strengths for single crystal Co$_{25}$Ni$_{25}$(HfTiZr)$_{50}$ micropillars with a variety of BCC and FCC metals and BMGs. The red dashed line shows the strength of bulk single crystal Co$_{25}$Ni$_{25}$(HfTiZr)$_{50}$. The size effect in Co$_{25}$Ni$_{25}$(HfTiZr)$_{50}$ is negligible. (**C**) A comparison of the mechanical loss factor (tan δ) versus

the elastic limit of conventional alloys, shape memory alloys, gum metal and the chemical complex $Co_{25}Ni_{25}(HfTiZr)_{50}$ alloy. (The data for shape memory alloys and gum metals are listed in Table S3 and S4). (**D**) The yield strength of single crystal $Co_{25}Ni_{25}(HfTiZr)_{50}$ micropillars at different temperature. Standard fits to the data (inset equation) yield a Peierls stress of $\tau_c \simeq \sigma_c /2.5 = 2.4$ GPa and effective temperature of $T_0 = 1107$ K. The inset shows the variation of the Young's modulus with temperature based on micro pillar compression tests. The data in **B** were obtained from measurements of 39 micropillars of different size, while the data in **D** came from 21 micropillars of the same diameter (~1μm).

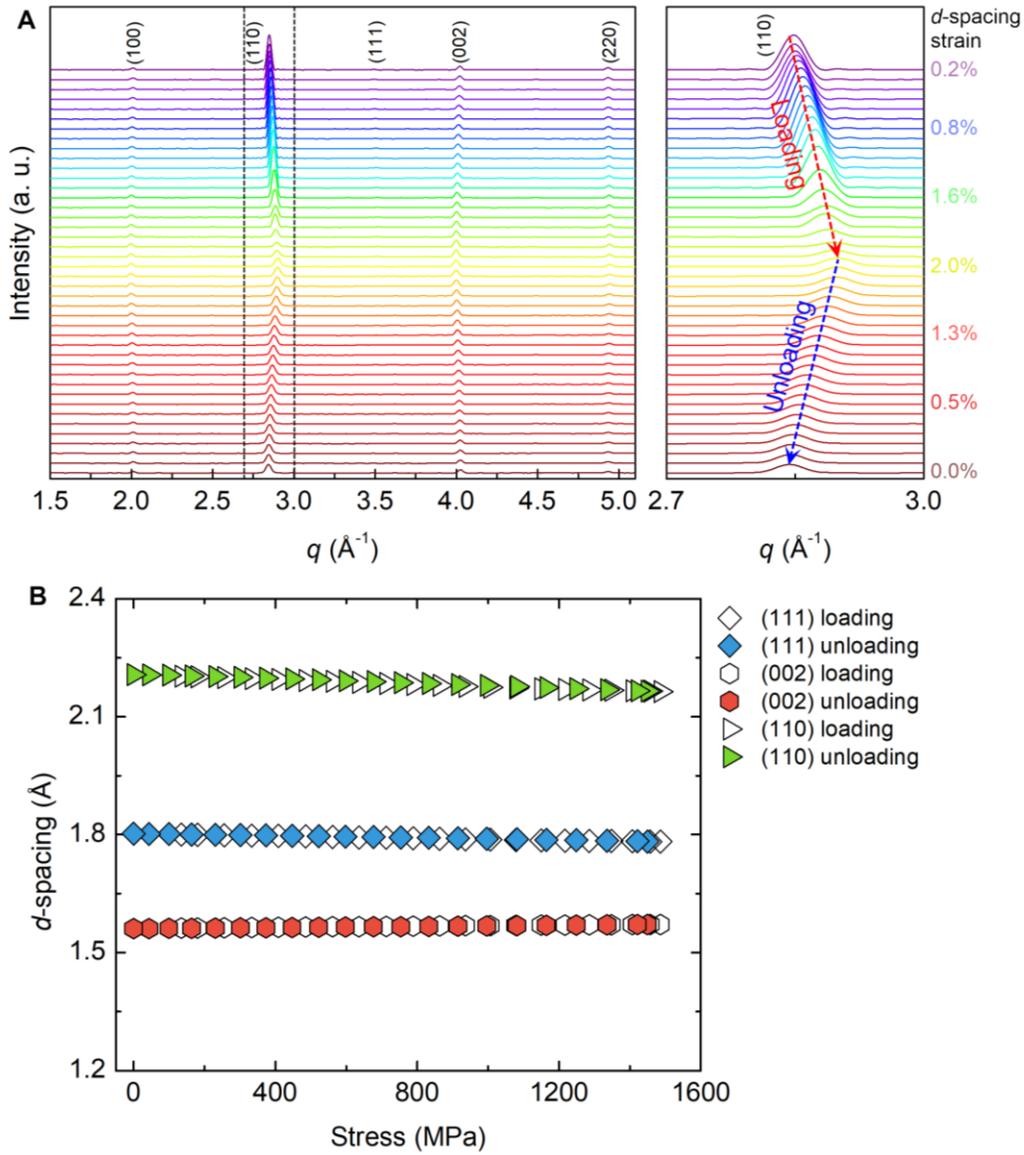

**Fig. 4. In situ high energy X-ray diffraction results for single crystal Co₂₅Ni₂₅(HfTiZr)₅₀ alloy.**
(**A**) The evaluation of diffraction peaks during the mechanical loading and unloading. The compression was conducted along the [110] direction. (**B**) The $d$-spacing evaluation during the mechanical loading and unloading process.

Interestingly, the inset of Fig. 3D shows that the $Co_{25}Ni_{25}(HfTiZr)_{50}$ alloy can keep to a constant elastic modulus as the temperature increased in the microcompression test, suggestive of an Elinvar effect (temperature independent elastic constants). To further corroborate the existence of a strong Elinvar effect, we measured the Young's modulus of our alloy using a resonance technique as a function of temperature (see **Methods** and Fig. S18). Fig. 5A demonstrates that, rather decreasing significantly with increasing temperature, the Young's modulus of the $Co_{25}Ni_{25}(HfTiZr)_{50}$ alloy increases slightly over the temperature range investigated (300 K to 900 K). This behavior is unique, and distinctly different from other structural alloys (*6*), including both BMGs and other medium entropy alloys (MEAs) (*44*). While the Elinvar effect is often attributed to magnetostriction and/or magnetoelastic effects (*45, 46*), the saturation magnetization of $Co_{25}Ni_{25}(HfTiZr)_{50}$ is very low (~1.17 emu/g at room temperature) (Fig. S19A). This alloy undergoes an antiferromagnetic to ferromagnetic phase transition at 851 K, which may explain the increase in the elastic modulus at ~ 900 K (Fig. S19B). Further examination shows that there is no magnetostriction effect in the $Co_{25}Ni_{25}(HfTiZr)_{50}$ alloy (Fig. S20). These results suggest that the strong Elinvar effect in $Co_{25}Ni_{25}(HfTiZr)_{50}$ likely does not have a magnetic origin.

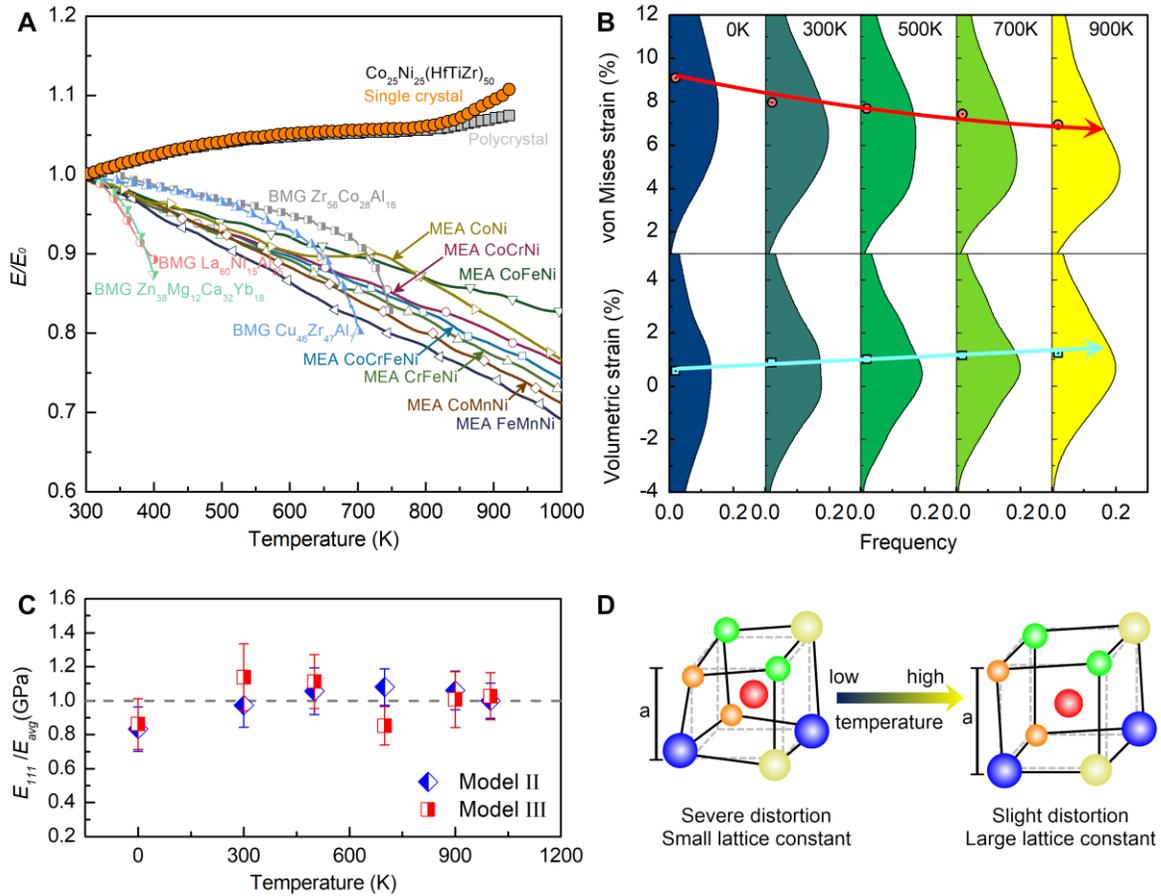

**Fig. 5. The outstanding enlinvar effect in the Co₂₅Ni₂₅(HfTiZr)₅₀ alloy.** (**A**) The Young's modulus versus temperature of the Co₂₅Ni₂₅(HfTiZr)₅₀ alloy compared with other metallic materials such as medium entropy alloys (*44*) and BMGs. $E_0$ is the young's modulus at room temperature and MEA stands for medium entropy alloy. (**B**) The distributions of von Mises and volumetric strain at different temperature of Model III structures obtained by the ab initio molecular dynamics simulations. The scatters are the average values of von Mises and volumetric strain. It shows that the average values of von Mises strain decrease while the average values of volumetric strain increase with the increasing temperature. (**C**) The elastic modulus variation with temperature of Co₂₅Ni₂₅(HfTiZr)₅₀ alloy obtained from the ab-initio molecular dynamics

simulations. $E_{[111]}$ is the modulus obtained from the uniaxial deformation along the [111] direction of Model II and III structure. (**D**) The schematic illustration of the variation of atomic structures with increasing temperatures.

We performed extensive ab initio molecular dynamics (AIMD) simulations to understand the distorted crystal structure and to identify the mechanism underlying the Elinvar effect in the $Co_{25}Ni_{25}(HfTiZr)_{50}$ alloy. Figure 5B shows that the average von Mises strain for Model III is ~9.09% at 0 K which decays, with increasing temperature, to ~6.94% at 900 K. On the other hand, the average volumetric strain in Model III increases from ~0.57% at 0 K to 1.25% at 900 K, indicative of a linear thermal expansion coefficient of ~$8.1 \times 10^{-6}$ $K^{-1}$ in this alloy (see Fig. S21; in agreement with our experimental data). Figure 5C shows that the elastic modulus $E_{111}$ (for Models II and III obtained from the AIMD) is remarkably temperature-independent, in excellent agreement with our experimental data (Fig. 5A); this is surprising in light of the fact that the elastic modulus of nearly all metallic alloys decreases with temperature.

Studies in amorphous materials suggest that elastic moduli will depend on both density and atomic packing topology (*47-49*). In nearly all materials, a decrease in density is accompanied with a decrease in elastic moduli; this occurs on heating in most materials (where the coefficient of thermal expansion is positive). In amorphous materials, increasing structural disorder tends to decrease the elastic moduli; this effect is even stronger than due to lattice expansion (*47*). The elastic modulus of as-cast amorphous alloys was observed to increase by 6%-30% while the corresponding density increased by only 0.5-2.0% upon the structural relaxation accompanying annealing (*48, 49*). In crystalline materials, structural disorder may be associated with point defects, chemical disorder, and other types of structural defects (*47-49*).

The AIMD and experimental data in Fig. 5 show that the $Co_{25}Ni_{25}(HfTiZr)_{50}$ volumetric strain increases on heating (positive thermal expansion coefficient), while the (von Mises) shear strain decreases on heating. The shear strain is a measure of disorder in the atomic packing. Hence, given the data presented in Fig. 5B, increasing temperature should decrease the elastic moduli (because of the positive coefficient of thermal expansion) but increasing temperature should simultaneously increase the elastic moduli (see Fig. 5B). In $Co_{25}Ni_{25}(HfTiZr)_{50}$, the two competing effects are nearly perfectly balanced; hence, the elastic moduli are nearly temperature independent (i.e., this is the origin of the Elinvar effect in our alloy). Note that the change in the lattice disorder with temperature is reversible with temperature cycling; it is a consequence of atomic vibration, not annealing-induced structural relaxation. These results suggest that engineering disorder provides a new route to creating temperature-independent, ultra-elastic behavior in a wide range of materials.

In conclusion, we demonstrate in this report that the $Co_{25}Ni_{25}(HfTiZr)_{50}$ alloy possesses a highly distorted lattice structure with a rather complex atomic-scale chemical order. Because of the combination of the unique structural features, our alloy attains a very high energy barrier against dislocation movements and defies the thermal softening as seen in conventional alloys. Consequently, our alloy displays a superb elastic strain limit, very low energy dissipation and an ultra-strong Elinvar effect, which is unmatched by conventional alloys as reported so far. This unique combination of elastic properties may find applications in many important fields, such as deep space exploration, in which people need high precision devices to function properly across a wide range of temperatures.

**Acknowledgements:**

The research of YY is supported by the Research Grant Council, the Hong Kong Government, through the General Research Fund (GRF) with the grant numbers CityU11213118 and CityU11200719 as well as by City University of Hong Kong with the grant number 9610391. CWP acknowledges the financial support of Academia Sinica Career Development Award with the grant number 2317-1050100. CWP is grateful for the computational support from the National Center for High-performance Computing, Taiwan. QFH is particularly grateful for the assistance given by Prof. X.K. Xi, Dr. X.D. Liu and Dr. T. Liu.

**Author contributions:**

Y.Y. supervised the project. Y.Y., C.W.P. and D.S. conceived the idea. Q.F.H. fabricated the polycrystalline samples and J.C.Q. prepared the single crystal samples. Q.F.H. characterized the structures and mechanical properties of the samples. J.G.W., C.W.P. and H.A.C. carried out the atomistic simulations. J.C.Q. and J.M.P. performed the dynamic mechanical spectroscopy analyses. C.T.L. performed the 3D APT experiments. L.H.X. and Q.S.Z. performed the in situ HEXRD experiments. Y.Y., D.S., C.W.P., Z.Y.D., Z.Q.Z. and Q.W. contributed to the data analysis. Y.Y., Q.F.H. and C.W.P. wrote the manuscript. All authors participated in the discussion of the results.

# Supplementary Materials

## Highly Distorted Lattices in Chemically Complex Alloys Produce Ultra-Elastic Materials with Extraordinary Elinvar Effects

Q.F. He[1†], J.G. Wang[1,2†], H.A. Chen[3†], Z.Y. Ding[1], Z.Q. Zhou[1], L.H. Xiong[4], J.H. Luan[5], J.M. Pelletier[6], J.C. Qiao[1,6,7], Q. Wang[8], L.L. Fan[9], Y. Ren[10], Q.S. Zeng[4,11], C.T. Liu[1,5], C.W. Pao[3*], D.J. Srolovitz[1,5*], Y. Yang[1,5*]

*Correspondence to: yonyang@cityu.edu.hk (YY); cwpao@gate.sinica.edu.tw (CWP) and srol@seas.upenn.edu (DS)

†Authors contribute equally to this work.

**Materials and Methods**

**Materials**

The polycrystalline samples of the chemical complex alloy $Co_{25}Ni_{25}(HfTiZr)_{50}$ (atomic %) were prepared through arc-melting in a high purity argon atmosphere. The purities of the raw materials for each element were at least 99.9 wt.%. The ingots were remelted at least four times to ensure the chemical homogeneity, and then suction cast into a copper mold. Two different types of copper mold (rod and plate) were used. The dimension of the cylindrical mold was 5 mm in diameter and 100 mm in length while that of the plate mold was $5 \times 10 \times 60$ mm$^3$. The single crystal $Co_{25}Ni_{25}(HfTiZr)_{50}$ alloys were prepared by a high rate directional solidification method following the standard procedure.

**Methods**

The microstructure characterization was performed on a scanning electron microscope (SEM, Quanta FEG 450). The crystalline orientation was determined by an electron backscattered diffraction (EBSD, EDAX TSL) equipped in a FEI Quanta FEG 250 SEM. The X-ray diffraction (XRD) instrument (Rigaku Smartlab) was used to identify the crystalline structure. The HADDF-STEM observations were conducted on JEM-ARM300F transmission electron microscope (TEM) equipped with double spherical aberration correctors. The identification of dislocation (Fig. S14-15) was performed on a JEOL 2100F TEM equipment. The TEM samples for STEM observation was prepared on a FEI scanning electron microscope/focused-ion beam (SEM/FIB) system following the normal ion-beam thinning procedure. The TEM samples for dislocation identification were prepared by mechanical grinding and ion-milling (PIPS II System, Gatan). The 3-D atom probe topography (APT) analysis was performed on the CAMECACA LEAP 4000X HR instrument. The APT samples were prepared in the SEM/FIB system using the lift-out method and annual milling.

The conventional or bulk compression tests were conducted on the MTS Alliance RT/30 Electro-Mechanical Materials Testing System with the strain rate of $1 \times 10^{-3}$ s$^{-1}$. The $2 \times 2 \times 4$ mm$^3$ compression samples were prepared by wire cutting, followed by mechanically grinding and polishing to a mirror finish. The effect of machine compliance effect was accounted for using the approach described in Ref. (1) using the Vit105 (Zr$_{52.5}$Ti$_5$Cu$_{17.9}$Ni$_{14.6}$Al$_{10}$ a/o) bulk metallic glass of a known Young's modulus (2) as the reference. Micropillar compression tests were conducted on a Hysitron TI 950 TriboIndenter systems equipped with a high temperature stage under an argon atmosphere. For the high temperature micropillar compression test, the sample was placed into a small chamber which was purged with pure argon continuously to avoid oxidation. There are two heaters at the bottom and top surface of the sample and the temperatures of the top and bottom heater plates were measured, in real-time, with two temperature sensors. The temperature difference of the top and bottom plates is ≤1°C. Micropillars with different top diameters were fabricated using the SEM/FIB systems. The height to diameter ratio of the micropillars was in the range of 2-3.

Dynamical mechanical analyses were performed in an inverted torsion pendulum in a high vacuum atmosphere following Ref.(*3*). The experiments were carried out by applying a sinusoidal stress at the fixed frequency of 0.3 Hz during continuous heating at the constant heating rate of 3 K/min. The testing samples had the dimension of $30 \times 3 \times 1.5$ mm$^3$. The complex modulus (G = G' +iG") was then deduced from the cyclic stress strain measurements, where G' represents the storage modulus while G" the loss modulus. The loss factor $\tan\delta = G^{"}/G^{'}$ could be then computed. The thermal expansion coefficient was measured on NETZSCH DIL 402C. Experiments were carried out with the constant heating rate 5 K/min and the sample size was $30 \times 3 \times 1.5$ mm$^3$. The Young's modulus measurement was conducted in the EG-UHT elastic modulus measurement system (NihonTechno-plus Co., Ltd) in the temperature range of 300-920 K. The sample plate was of $50 \times 3 \times 1.3$ mm$^3$. The measurement was carried out in the bending mode. The Young's modulus was determined from the resonant frequency $f_0$ ( $E \sim f_0^2$ ). The magnetization curve was measured by a vibrating sample magnetometer (VSM). The temperature dependence of magnetization was measured by a magnetic property measurement system (MPSM). The magnetostriction behaviour was studied in a pulsed magnetic field. The pulsed magnetic field is produced by discharge of the pulse magnet which consists of a bank of charged capacitors ($\sim$ 6800 $\mu$F) and a drive coil ($\sim$ 15 $\mu$H) with length and diameter of L$\times$D=100$\times$100 mm$^2$. The test equipment includes four components, i.e., a strain gauge as strain sensor in an ac bridge, a lock-in amplifier for strain signal conditioning, an oscilloscope for data acquiring and recording, and a personal computer for data processing. The resolution of the magnetostriction coefficient measurement setup is 2ppm.

In-situ high-energy x-ray diffraction (HEXRD) compression test was conducted at the 11-ID-C beamline in the Advanced Photon Source (APS), Argonne National Laboratory, by using a monochromatic x-ray beam with a wavelength of 0.1173 Å and beam size of 0.5 mm × 0.5 mm. Two-dimensional diffraction patterns in the transmission geometry were obtained by using a Perkin-Elmer large area detector downstream set at 1.7 m away from the sample. Gaussian fits were used to resolve diffraction peak position and areas and error of the d-spacing strain measurement was estimated to be smaller than 0.1%. Sub-sized compression samples with the dimension of 1.8 mm × 1.8 mm × 3.9 mm were mounted in a custom-built loading frame and the in-situ compression tests were performed at a strain rate of $5 \times 10^{-3}$ s$^{-1}$.

Structural relaxations and energy calculations were performed using Vienna Ab initio Simulation Package (VASP)(*4-6*). The generalized gradient approximation (GGA)(*7, 8*) was used with the Perdew-Burke-Ernzerhof (PBE) exchange-correlation functional and the projector augmented wave (PAW) pseudopotentials(*9, 10*). The cutoff energy was 400 eV and a 3×3×3 Monkhorst-Pack k-point mesh was used in all calculations. The convergence criteria of self-consistent field for structural relaxations were set to be $10^{-5}$ eV for the total energy and $10^{-2}$ eV/Å for the atomic force in structure optimizations.

To study the local atomic scale lattice structures, we constructed three types of $Co_{25}Ni_{25}(HfTiZr)_{50}$ crystal structure model for structural relaxation calculations, including (i) Model I: chemically disordered BCC with atoms being randomly assigned to lattice sites, (ii) Model II: the corner and center sites of a BCC unit cell are respectively occupied by the {Co, Ni} and {Hf, Ti, Zr} atoms, and (iii) Model III: partially ordered BCC in line with our STEM observations, in which 25% of Zr atoms exchange lattice sites with Ni and Co atoms for Model II.

In order to compute the local atomic strains in the structural models, the corresponding lattices were constructed at 0 K. At the first, only lattice expansion and contraction were allowed during structural relaxation. The lattice structure so determined had a local energy minimum and was considered to be the reference lattice. After that, further structural relaxations were performed at different temperatures, which allowed atoms to displace from their ideal positions in addition to lattice expansion/contraction. The strain fields in these further relaxed lattices with respect to the reference lattice were derived based on the Green-Lagrangian strain tensor (*11, 12*) (denoted as $\varepsilon$ hereafter). Two types of atomic strains, including the volumetric strain $\varepsilon_v$ and the von Mises strain $\gamma$, were calculated. The volumetric strain $\varepsilon_v$ is defined as the mean of the three normal strains: $\varepsilon_v = \left( \varepsilon_{xx} + \varepsilon_{yy} + \varepsilon_{zz} \right) / 3$. The von Mises strain, also known as the von Mises local shear invariant, is defined as $\gamma = \sqrt{\varepsilon_{xy}^2 + \varepsilon_{yz}^2 + \varepsilon_{zx}^2 + \frac{1}{6}\left[ \left( \varepsilon_{xx} - \varepsilon_{yy} \right)^2 + \left( \varepsilon_{yy} - \varepsilon_{zz} \right)^2 + \left( \varepsilon_{zz} - \varepsilon_{xx} \right)^2 \right]}$. To calculate the Young's modulus along [111] in Model II and III through the ab initio simulations at different temperatures, the structural models were first relaxed under the zero pressure and then were then deformed uniaxially along the z direction, which was oriented along [111]. The strained structures were relaxed to reach an energy minimum. The potential energy F is a function of uniaxial strain $\varepsilon$, i.e. $F = F_0 + V_0 \sigma \varepsilon + \frac{V_0}{2} E^* \varepsilon^2$, where the effective Young's modulus can be obtained through data fitting (*13, 14*). In theory, it can be easily derived that is $E^* = C_{11} + 2C_{12}$ or $E^* = \beta \cdot E_{111}$, where $\beta = \frac{\mu}{3\mu - E_{111}}$ for a cubic system and $\beta = \frac{1}{1 - 2\upsilon}$ for an isotropic system, in which $\mu$ is the shear modulus and $\upsilon$ is poisson's ratio. Assuming $\upsilon$=0.25 (typical of BCC metals (*15*)), we can obtain $E_{111} = 91.3$ GPa and 87.7 GPa for Models II and III, respectively.

## Determination of atomic size difference

The atomic size difference for multicomponent alloys can be quantified as $Delta = \sqrt{\sum_{i=1}^{N} c_i \left(1 - r_i / \bar{r}\right)^2}$ where $N$ is the number of constituent elements in the alloy, $c_i$ the atomic fraction of the *i-th* component, $r_i$ the atomic radius of the *i-th* component and $\bar{r}$ the average atomic radius. For the present $Co_{25}Ni_{25}(HfTiZr)_{50}$ alloy, the atomic radii of the constituent elements are 1.251 Å for Co, 1.246 Å for Ni, 1.578 Å for Hf, 1.462 Å for Ti and 1.603 Å for Zr (*16*). The calculated atomic size difference of the $Co_{25}Ni_{25}(HfTiZr)_{50}$ is about 11%.

## Weibull statistics analysis

In particular, the Weibull statistics can be studied through the following three-parameter equation (*17-19*):

$$P_f = 1 - exp\left[-D^p \left(\frac{\sigma - \sigma_u}{\sigma_0}\right)^n\right] \qquad \text{(Eq. S1)}$$

where $P_f$ is the probability of yielding or failure, $\sigma$ is an applied uniaxial stress; $\sigma_0$ is a reference stress, $n$ is the Weibull modulus, $p$ is dimensionality (2 for surface and 3 for volume) and $D$ is the characteristic dimension of the samples. The parameter $\sigma_u$ is the stress at which the failure probability is zero; it is usually set to zero. According to (*17*), the Weibull modulus $n$ was calculated to be about 27 for our alloy, the Weibull modulus $n$ was calculated to be about 27 for our alloy, based on our micropillar compression (Fig. 3B) and bulk compression (Fig. 3A) data. Following Ye et al. (*20*), this corresponds to $m \sim -0.07$ in the empirical rule $\frac{\sigma}{E} = k \cdot D^m$; we used this for data fitting for surface initiated yielding. Because $m$ is so small (-0.07) compared to that for typical FCC ($m \approx -0.6$) and BCC ($m \approx -0.35$) metals, the effect of sample size on the yield strength of our $Co_{25}Ni_{25}(HfTiZr)_{50}$ single crystals is very small. On the other hand, if we extrapolate the size trend from the micro-compression data to the bulk specimen size, one clearly sees that the bulk specimen data is well-described (within the range of the 95% confidence fitting curves) – see Fig. S10

**Determination of activation volume from the stress relaxation test**

The stress relaxation tests were conducted in the Hysitron TI950 TriboIndenter system. Fig. S16 shows one typical curve of stress versus time. The micropillar was first deformed at a constant strain rate to 10% total strain. After that, the total displacement was kept at a constant and the micropillar underwent continuous stress relaxation. According to Ref.(*21*), the stress decays in a logarithmic manner:

$$\Delta \sigma = -\frac{kT}{MV} \ln\left(1 + t/c_r\right)$$

(Eq. S2)

where $\Delta \sigma$ is the value of stress drop, $k$ the Boltamann constant, $T$ the temperature, $M$ the Schmid factor which is about 0.47 in this study, $V$ the activation volume, $t$ the relaxation time and $c_r$ is a time constant which can be obtained from data fitting. In our study, the activation volume can be obtained by fitting the curve in Fig. S16 to Eq. S2, which gives $V$=3.05±0.68b$^3$.

**Stochastic Peierls-Nabarro (PN) model**

To rationalize the high Peierls stress of ~2.4 GPa for the present Co$_{25}$Ni$_{25}$(HfTiZr)$_{50}$ alloy, we resort to the stochastic Peierls-Nabarro model developed very recently by Zhang et al.(*22*). In this model, chemical randomness and chemical short-range order (CSRO) are both considered to perturb the Peierls potential for an enhanced Peierls stress (*22*). As such, the Peierls stress of a chemically complex alloy can be derived as follows:

$$\tilde{\tau}_c \approx 1 + \frac{8z_0\zeta_0^2}{b\sqrt{\lambda}} \Delta \sqrt{\int_{-\infty}^{+\infty}\left[\int_s^{+\infty} \exp\left(-\frac{x-s}{\lambda}\right)\frac{x^2}{\left(x^2+\zeta_0^2\right)^2} dx\right]^2 ds}$$

(Eq. S3)

Where $\lambda$ is the correlation length characteristic of the CSRO; $\tilde{\tau}_c = \tau_c / \tau_p^0$ in which $\tau_c$ is the critical stress increment associated with the local atomic scale randomness; $\tau_p^0$ is the reference Peierls stress that can be estimated via the simple rule of mixture; and $z_0 = erfc^{-1}\left(2\tau_p^0 / \mu\right)$ in which $\mu$ is the shear modulus, $\zeta_0$ the dislocation core size; and $b$ is the magnitude of the Burger's vector.

For the $Co_{25}Ni_{25}(HfTiZr)_{50}$ alloy, the magnitude of the Burger's vector $b$ is about 3.12 Å obtained from in-situ high-energy x-ray diffraction; the shear modulus $\mu$ is estimated about 40.8GPa; and the reference Peierls stress $\tau_p^0$ can be estimated as $\tau_p^0 = c_i\left(\tau_p^0\right)_i \approx 54.2 MPa$, where the Peierls stress $\left(\tau_p^0\right)_i$ of the constituent elements Co, Ni, Ti, Hf and Zr are 0.3MPa, 5.54MPa, 48.17MPa, 48.8MPa and 8.6MPa, respectively(*23, 24*). By taking $\zeta_0 \approx b$, we can compute $\tau_c$ of the $Co_{25}Ni_{25}(HfTiZr)_{50}$ alloy, which is a function of $\Delta$ and $\lambda/\zeta_0$. As seen in Fig. S17, the exceptionally high Peierls stress ($\tau_c \approx 2.4$ GPa) corresponds to a large $\lambda/\zeta_0$ ratio and a high $\Delta$, or a significantly perturbed Peierls potential. On a qualitative basis, this is in line with the highly distorted lattice, strong atomic scale chemical ordering and overall chemical randomness, as seen in our alloy.

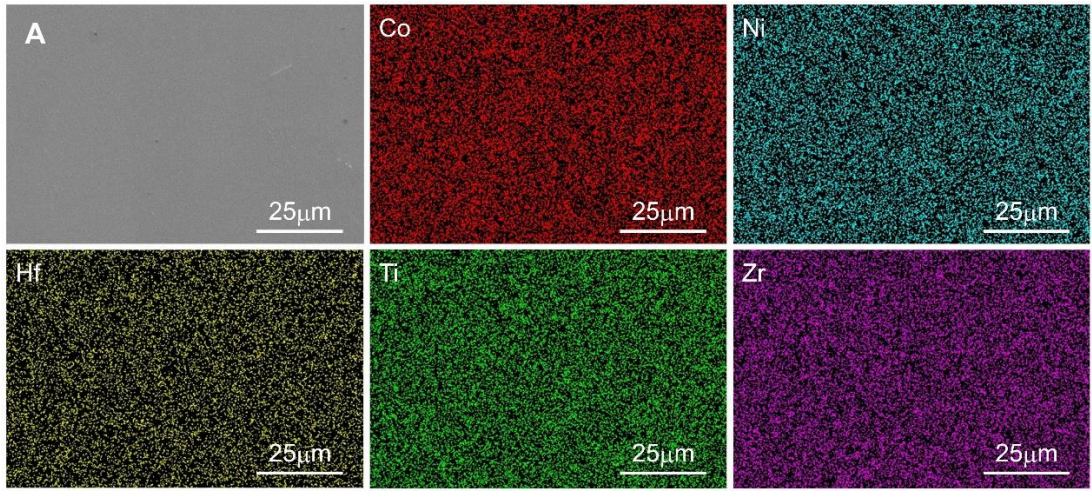

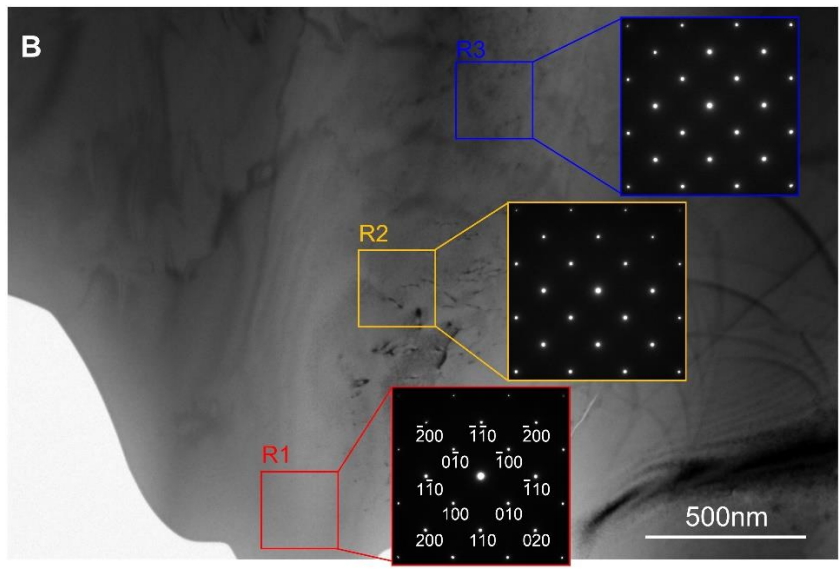

**Fig. S1.**

**Characterization of single crystal Co₂₅Ni₂₅(HfTiZr)₅₀ alloy.** (**A**) A low magnification backscatter electron image and elemental distributions show that the chemical distribution is homogeneous in the single crystal samples on a sub-micro scale. (**B**) Low magnification TEM image and the corresponding diffraction patterns ([001] zone axis) in different regions show that there is no phase separation in the single crystal samples.

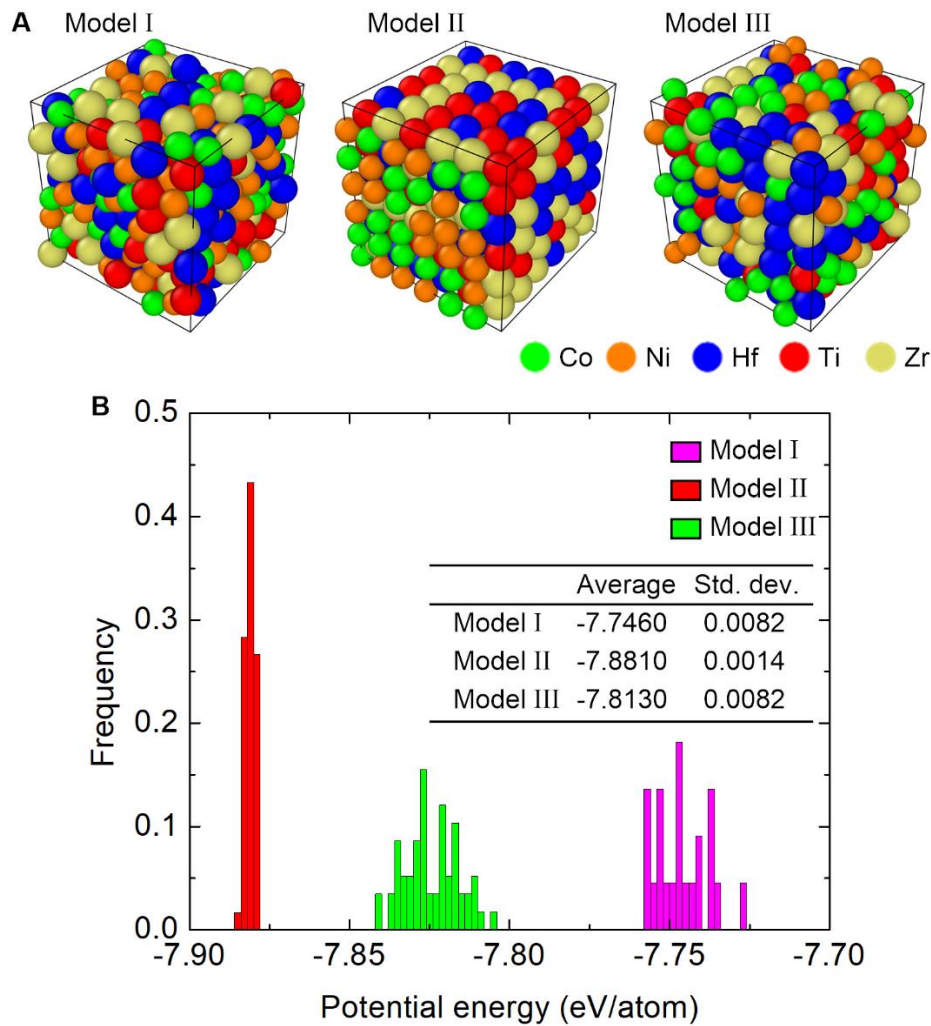

**Fig. S2.**

**The calculated potential energies of the three structure models for Co₂₅Ni₂₅(HfTiZr)₅₀ at 0K.**

(**A**) The typical atomic structures for the three structure models. (**B**) The comparison of the distributions of the calculated potential energy of the three structure models.

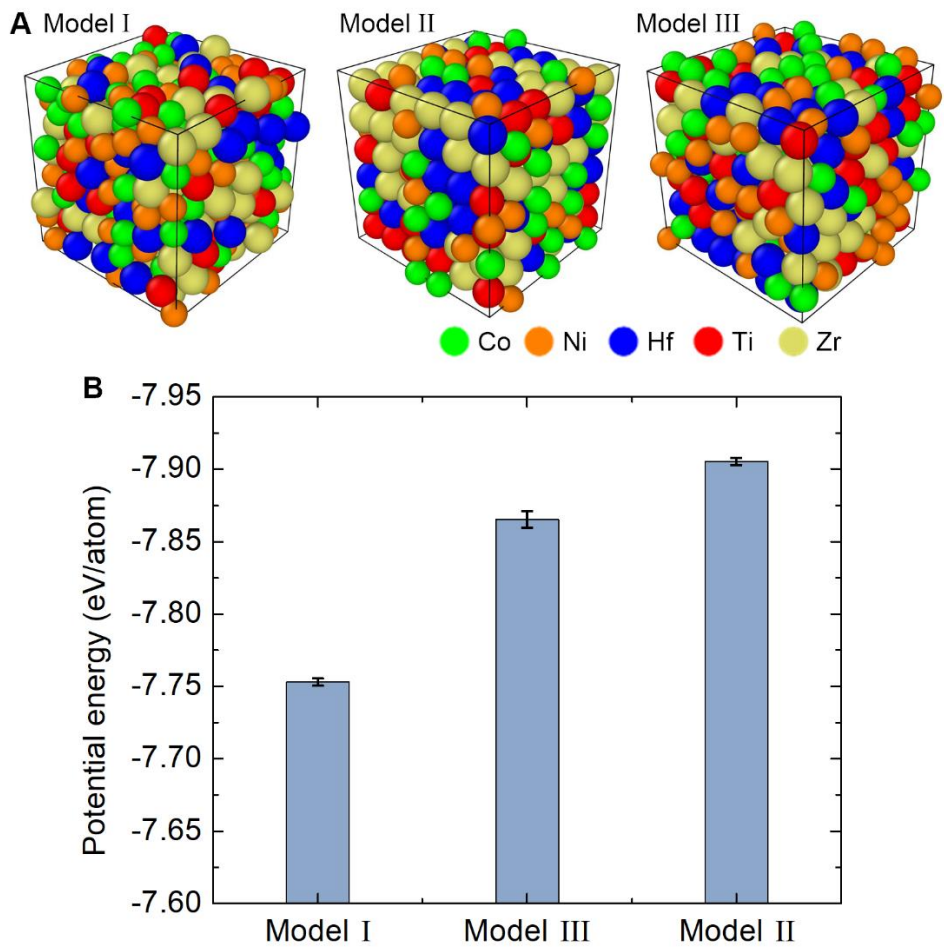

**Fig. S3.**

**The calculated potential energies of the three structure models for Co$_{25}$Ni$_{25}$(HfTiZr)$_{50}$ at 300K.** (**A**) The typical atomic structures for the three structure models. (**B**) The comparison of the calculated potential energy of the three structure models.

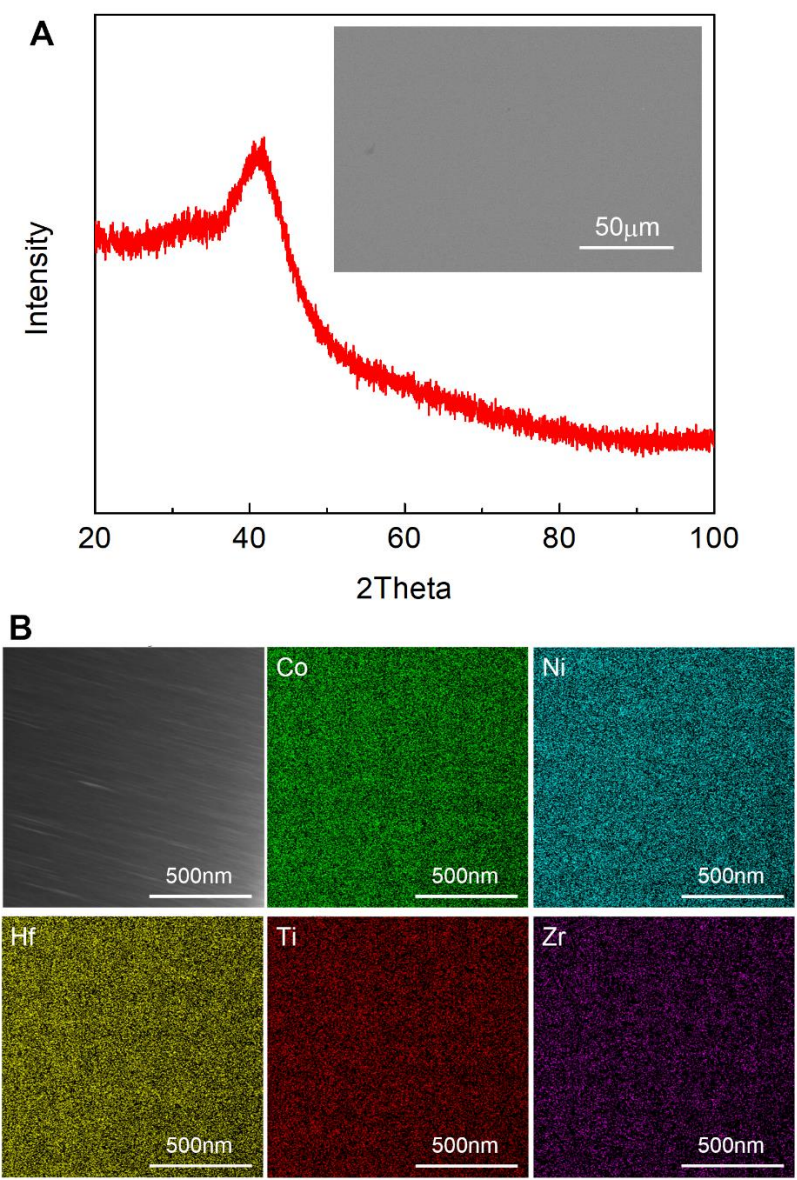

**Fig. S4.**

**The structural characterization of Co₂₅Ni₂₅(HfTiZr)₅₀ thin film.** (**A**) The XRD spectrum of the Co₂₅Ni₂₅(HfTiZr)₅₀ magnetron sputtered film. The inset SEM image shows the surface microstructure of the Co₂₅Ni₂₅(HfTiZr)₅₀ thin film. (**B**) The STEM image and elemental distributions of Co₂₅Ni₂₅(HfTiZr)₅₀ thin film.

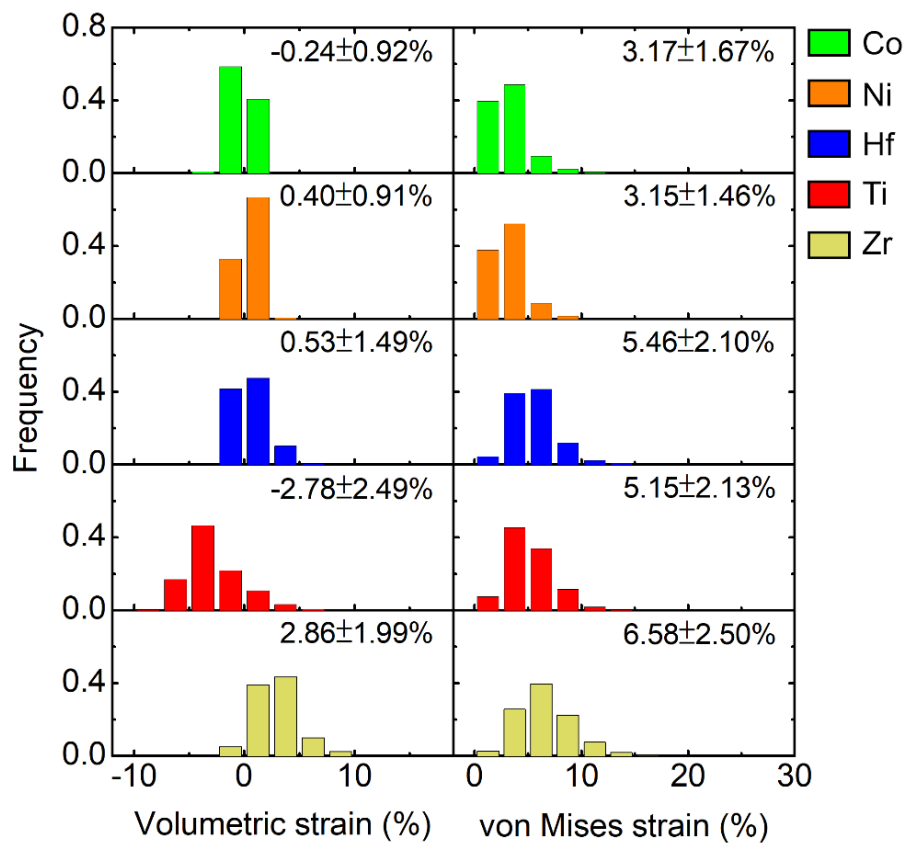

**Fig. S5.**

The calculated distribution of the atomic-scale volumetric and von Mises strain of Model II.

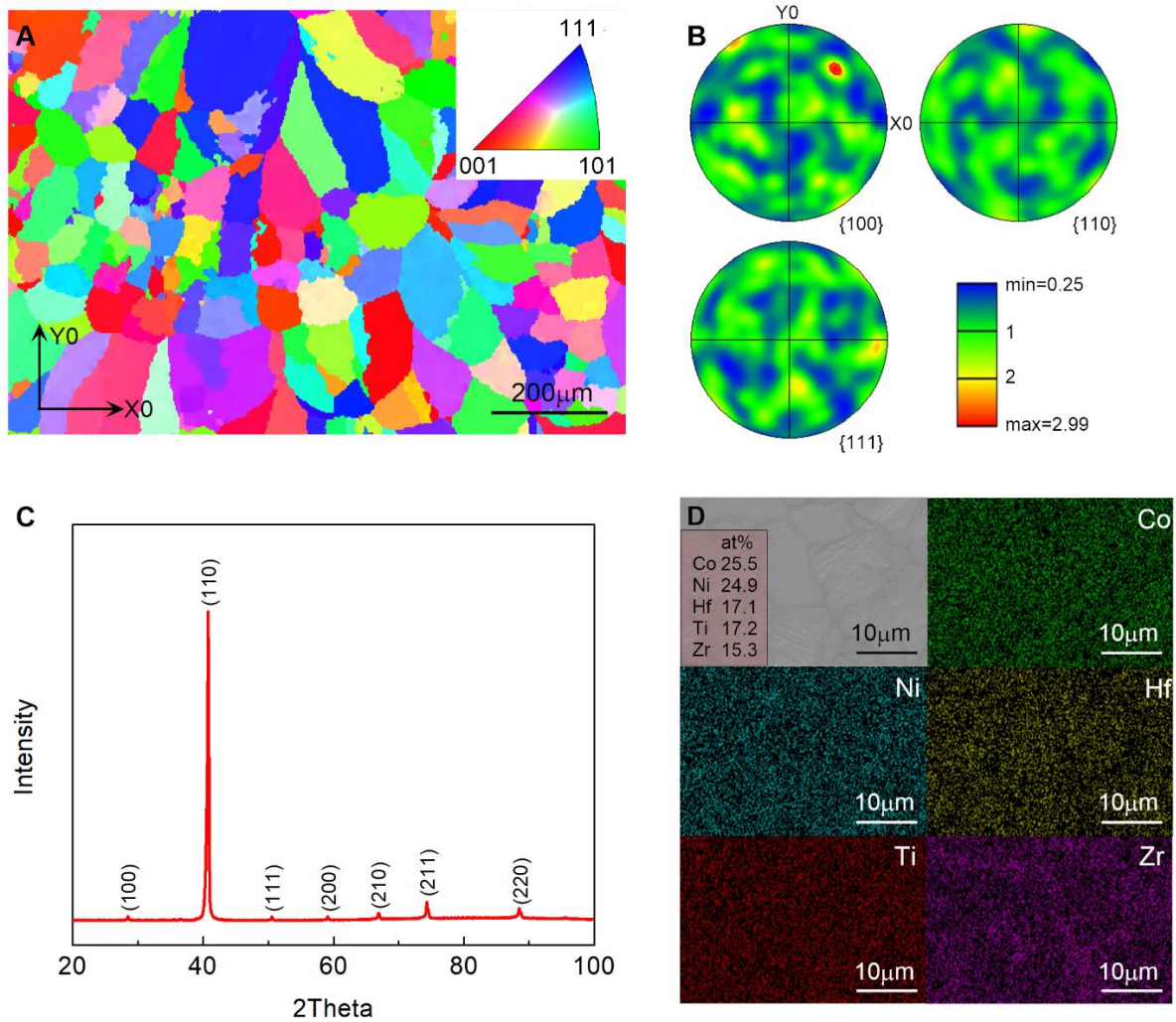

**Fig. S6.**

**The structure characterization of the as-cast polycrystalline Co₂₅Ni₂₅(HfTiZr)₅₀.** (**A**) The inverse pole figure (IPF) map showing the grain structure of as-cast polycrystalline Co₂₅Ni₂₅(HfTiZr)₅₀ alloy. (**B**) The pole figures showing that there are no preferred orientations in the as cast polycrystalline Co₂₅Ni₂₅(HfTiZr)₅₀ alloy. (**C**) The XRD pattern of the polycrystalline Co₂₅Ni₂₅(HfTiZr)₅₀ indicative of a single phase B2 structure. (**D**) The SEM image showing the microstructure of the as-cast polycrystalline Co₂₅Ni₂₅(HfTiZr)₅₀ sample and the corresponding elemental mappings.

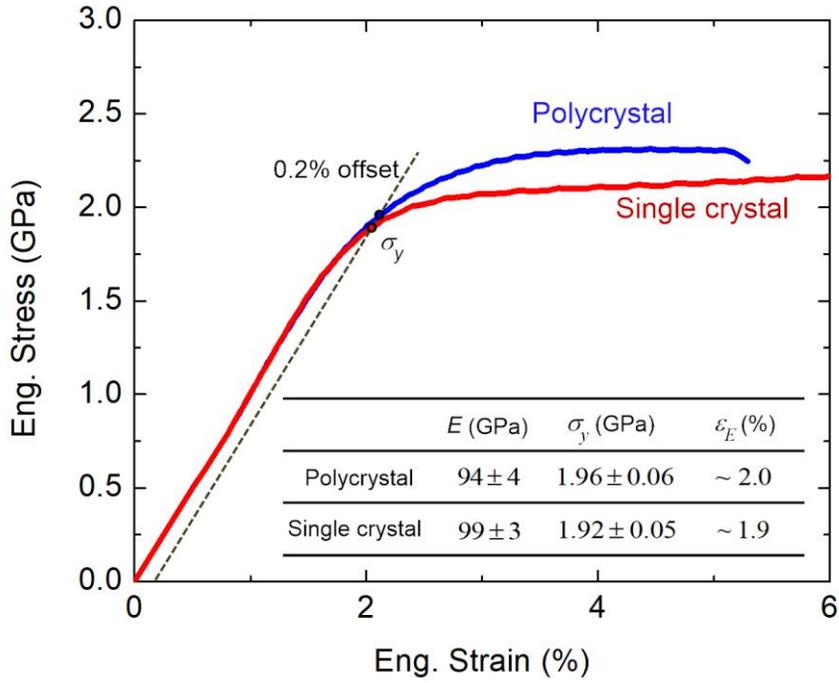

**Fig. S7.**

**Representative stress-strain curves from bulk compression (uniaxial) testing of polycrystalline and single crystal Co$_{25}$Ni$_{25}$(HfTiZr)$_{50}$ alloy samples.** The Young's modulus was obtained through linear fitting of elastic portion of the stress-strain curves. The yield strengths were determined as the amount of stress at 0.2% offset yield strain (black dash line). The elastic strain limit was determined from the relation of $\varepsilon_E = \sigma_y/E$. The values of the Young's modulus $E$, yield strength $\sigma_y$ and elastic strain limit $\varepsilon_E$ are averages from three compression tests.

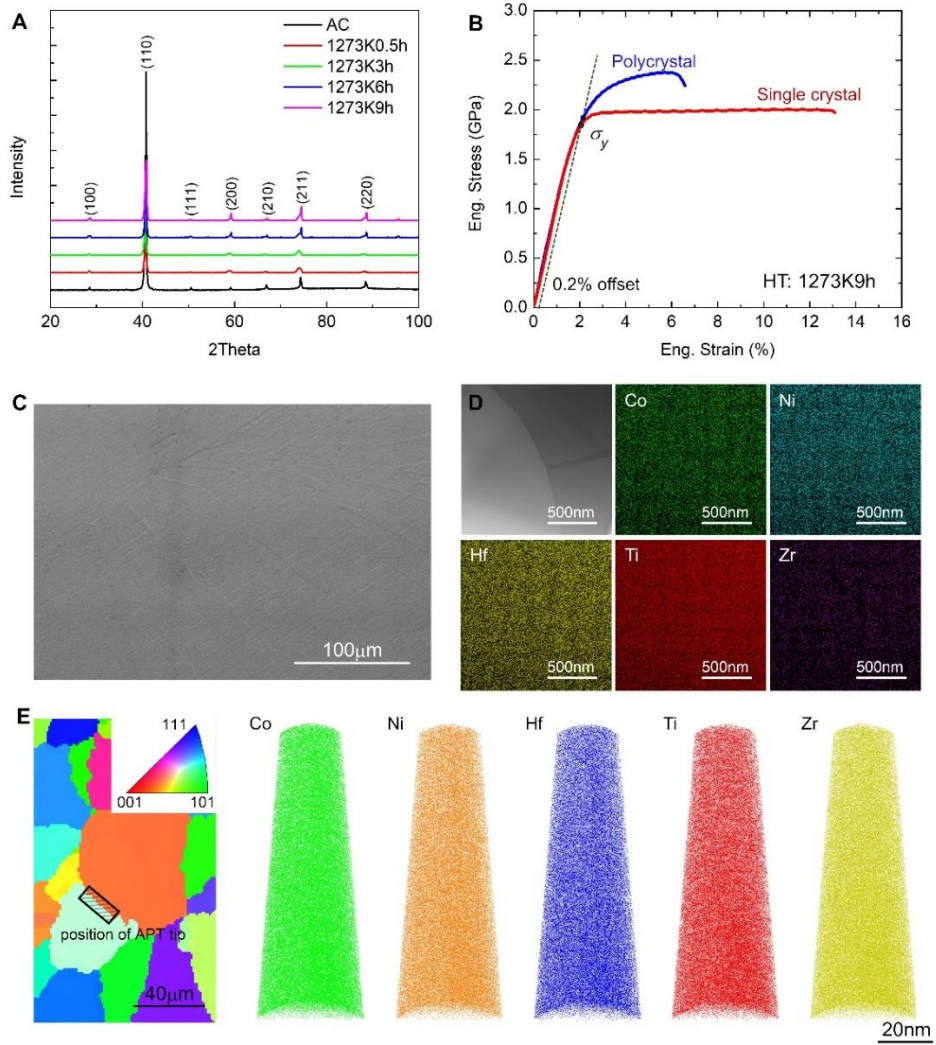

**Fig. S8.**

**Structure and mechanical property characterization of the Co$_{25}$Ni$_{25}$(HfTiZr)$_{50}$ alloy after thermal annealing.** (**A**) XRD patterns of the Co$_{25}$Ni$_{25}$(HfTiZr)$_{50}$ alloy sample after thermal annealing at 1273K for different times. It shows that the Co$_{25}$Ni$_{25}$(HfTiZr)$_{50}$ alloy remains a single phase, B2 structure under all annealing conditions examined. (**B**) Compression stress-strain curves of the single and polycrystal Co$_{25}$Ni$_{25}$(HfTiZr)$_{50}$ alloy after annealing at 1273K for 9 hours. The results show that the mechanical properties of the Co$_{25}$Ni$_{25}$(HfTiZr)$_{50}$ alloy do not change after the heat treatment. (**C**) A low magnification SEM image shows the microstructure of the Co$_{25}$Ni$_{25}$(HfTiZr)$_{50}$ alloy after annealing at 1273K for 9 hours. (**D**) The STEM image and elemental mapping near a grain boundary. No segregation to GBs was observed following a 9 hour, 1273K anneal in our Co$_{25}$Ni$_{25}$(HfTiZr)$_{50}$ alloy. (AC and HT represent as-cast and heat treated respectively.) (**E**) The APT reconstructions of the three-dimensional elemental distributions showing the chemical homogeneity along the grain boundary at the nanometer scale. The black rectangle in the EBSD image indicates the position from which the APT tip was carved.

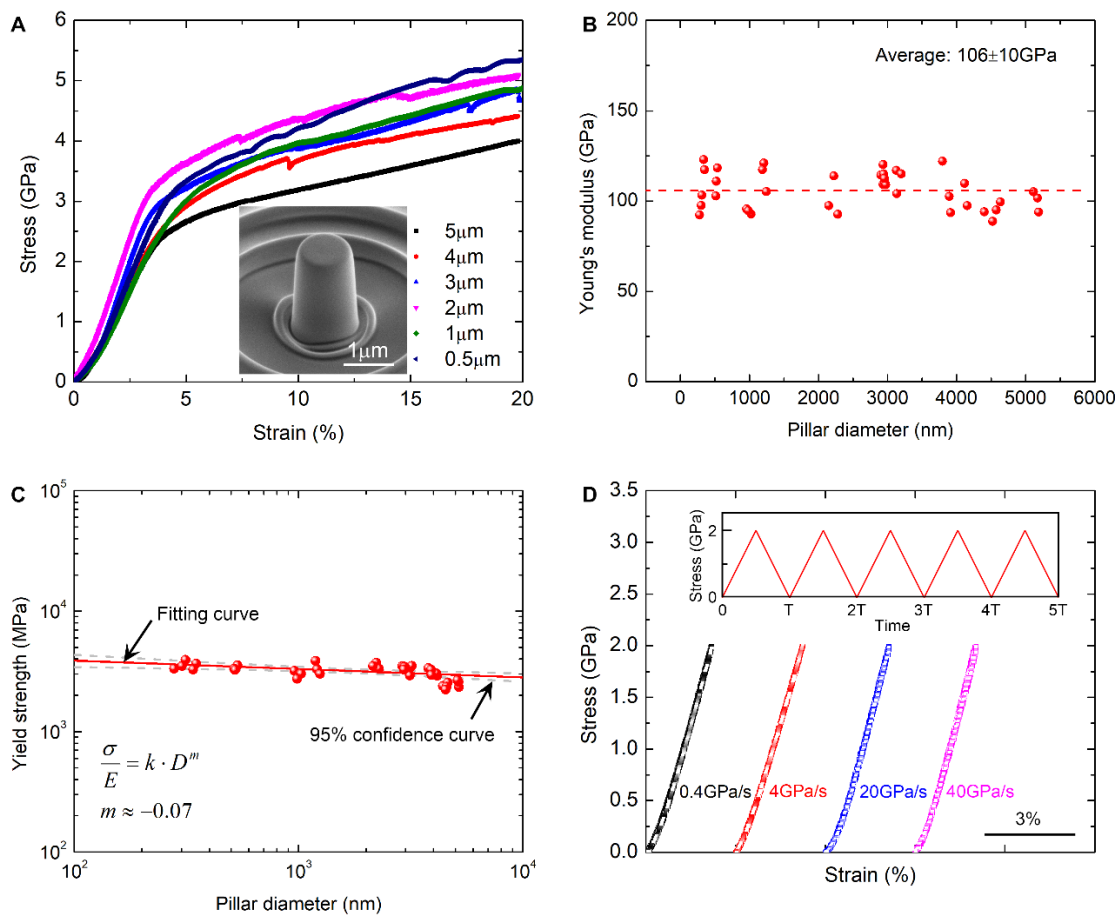

**Fig. S9.**

**The monotonic and cyclic microcompression results of the single crystalline Co$_{25}$Ni$_{25}$(HfTiZr)$_{50}$.** (**A**) The typical monotonic stress-strain curves obtained for different micropillar diameters. The inset shows a typical pillar image. (**B**) The Young's modulus versus pillar diameter of these micro pillars in Fig. 3B. The average Young's modulus is measured to be 106 GPa following the method detailed in (*25*). (**C**) The size dependent yield strength of single crystal Co$_{25}$Ni$_{25}$(HfTiZr)$_{50}$ micro pillars. (**D**) The cyclic stress-strain curves obtained within the elastic regime from the micropillar at different nominal stress rates. Note that no mechanical hysteresis is evident in (**D**). The inset shows the (cyclic) load versus time. Five cycles were performed for each compression. The geometrical dimensions of micropillars used in this study are listed in **Supplementary Table S5**.

**Fig. S10.**

**The dependent strengths of micro pillar and bulk compression sample for single crystal Co$_{25}$Ni$_{25}$(HfTiZr)$_{50}$.**

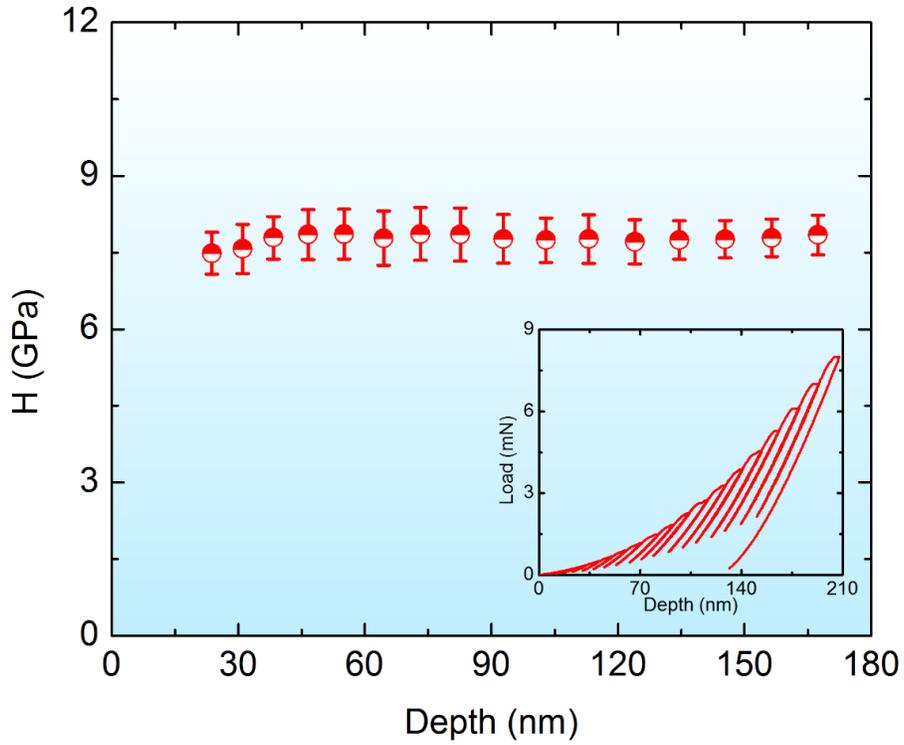

**Fig. S11.**

The hardness as a function of indentation depth measured through nanoindentation on the single crystalline $Co_{25}Ni_{25}(HfTiZr)_{50}$. Note that the experimental results clearly show no indentation size effect. The inset is the cyclic loading curve comprising different loading and partial unloading segments.

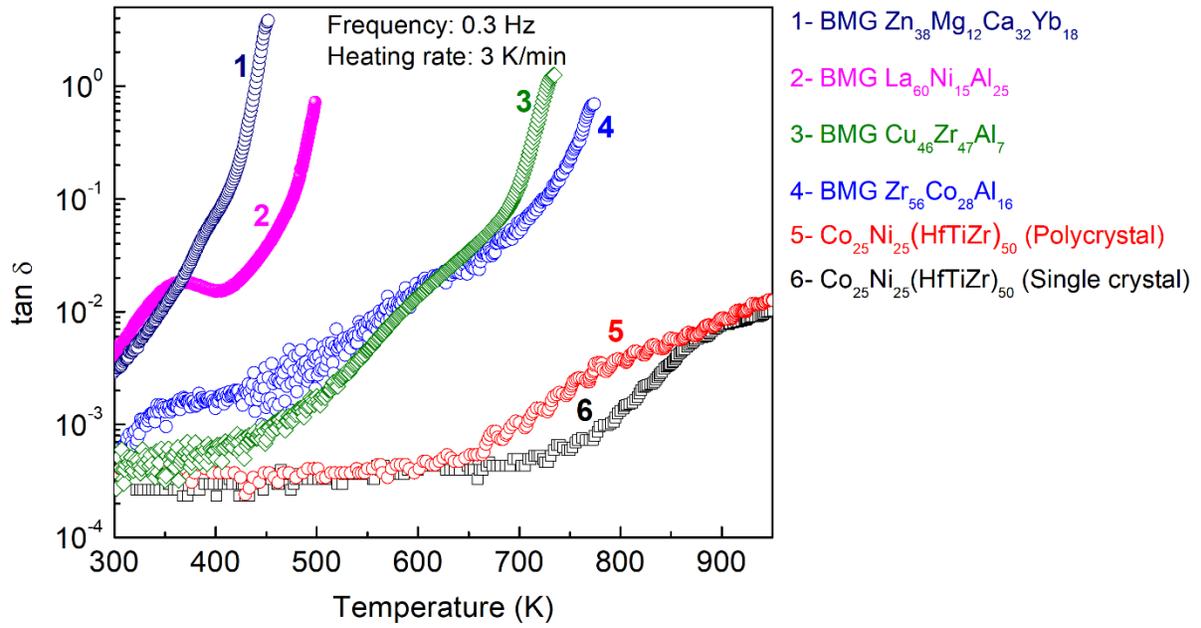

**Fig. S12.**

The loss factors measured for the single- and poly-crystalline $Co_{25}Ni_{25}(HfTiZr)_{50}$ in comparison with various bulk metallic glasses over a wide temperature range.

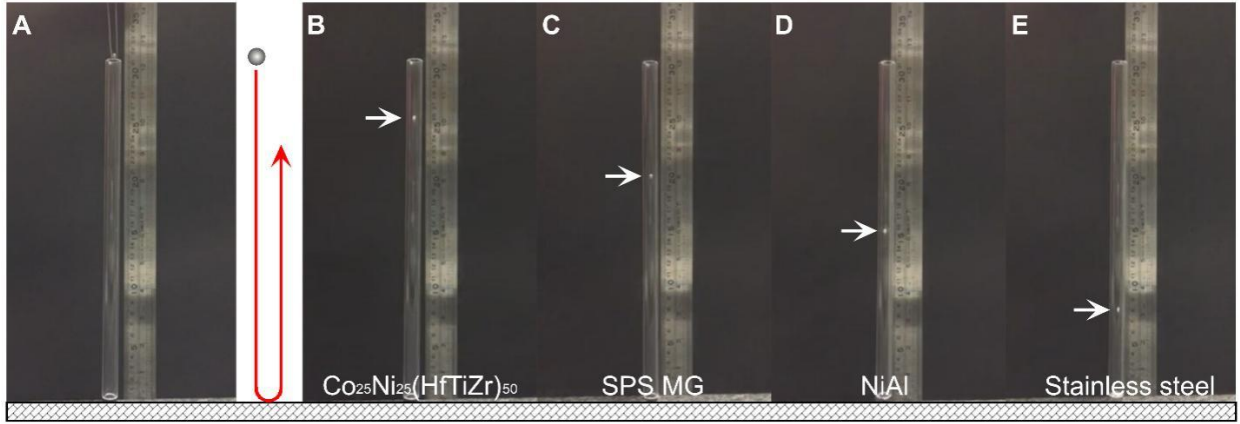

**Fig. S13.**

**Comparison of the first height of the steel ball bouncing back from different alloy surfaces.**

(**A**) The photo showing the starting moment of the bouncing experiments. The first height of the steel ball after bouncing back, as indicated by the white arrow, from (**B**) the single crystalline $Co_{25}Ni_{25}(HfTiZr)_{50}$, (**C**) the spark plasma sintered (SPS) $Cu_{50}Zr_{45}Al_5$ metallic glass, (**D**) the NiAl alloy with a B2 structure and (**E**) the commercial stainless steel. Please refer to **Movie S1** for details. Note that all the bulk alloys had the similar size.

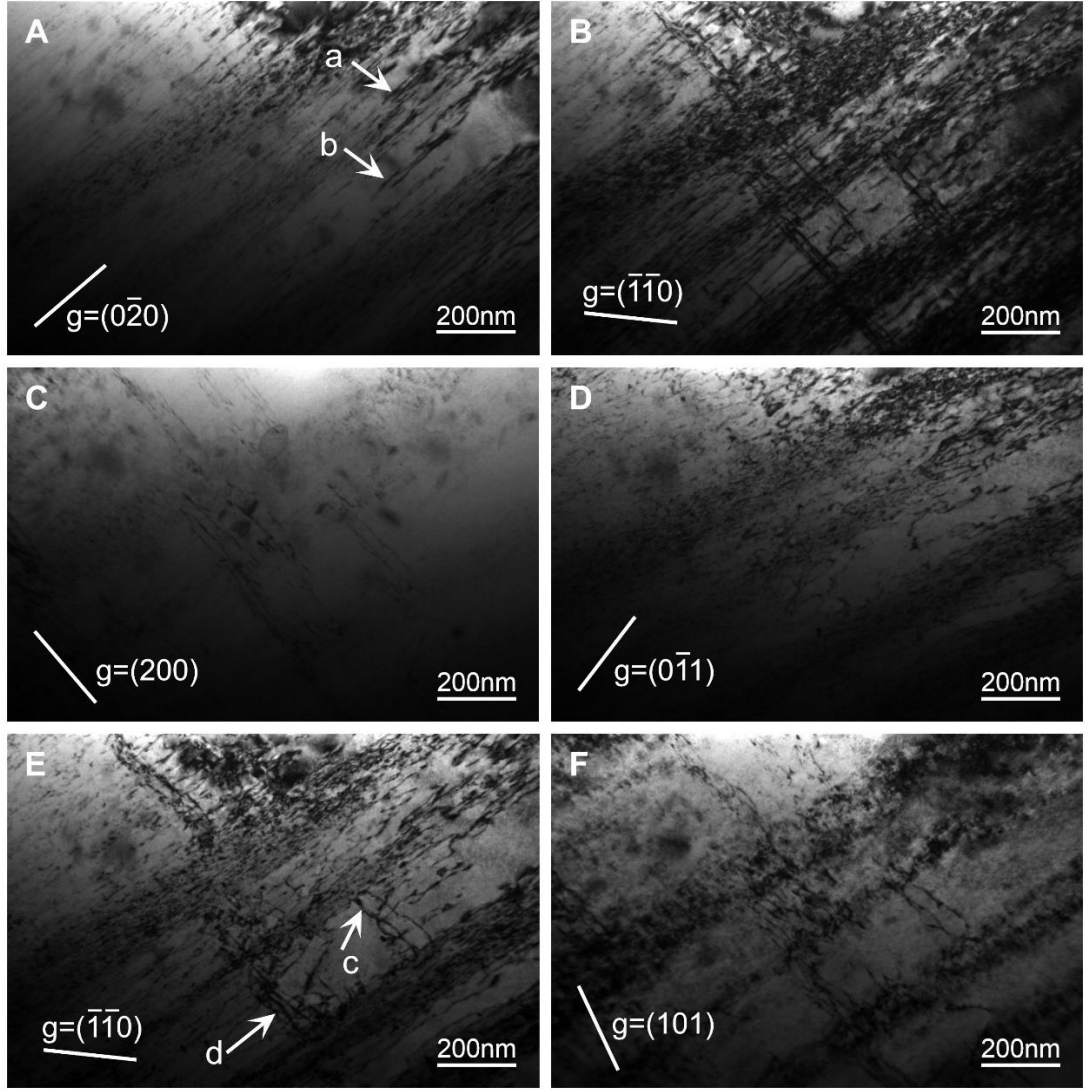

**Fig. S14.**

**The dislocation structure analysis performed in the [111] single crystalline Co$_{25}$Ni$_{25}$(HfTiZr)$_{50}$ sample after deforming to 4% mechanical strain.** (**A**) $g = (0\bar{2}0)$ and the beam direction $Z = [001]$, (**B**) $g = (\bar{1}\bar{1}0)$ and the beam direction $Z = [001]$, (**C**) $g = (200)$ and the beam direction $Z = [001]$, (**D**) $g = (0\bar{1}1)$ and the beam direction $Z = [\bar{1}11]$, (**E**) $g = (\bar{1}\bar{1}0)$ and the beam direction $Z = [\bar{1}11]$ and (**F**) $g = (101)$ and the beam direction $Z = [\bar{1}11]$. The **g•b**=0 out of contrast analyses indicate that the dislocations are of the <001> type. Please see **Supplementary Table S1** for the detailed analyses.

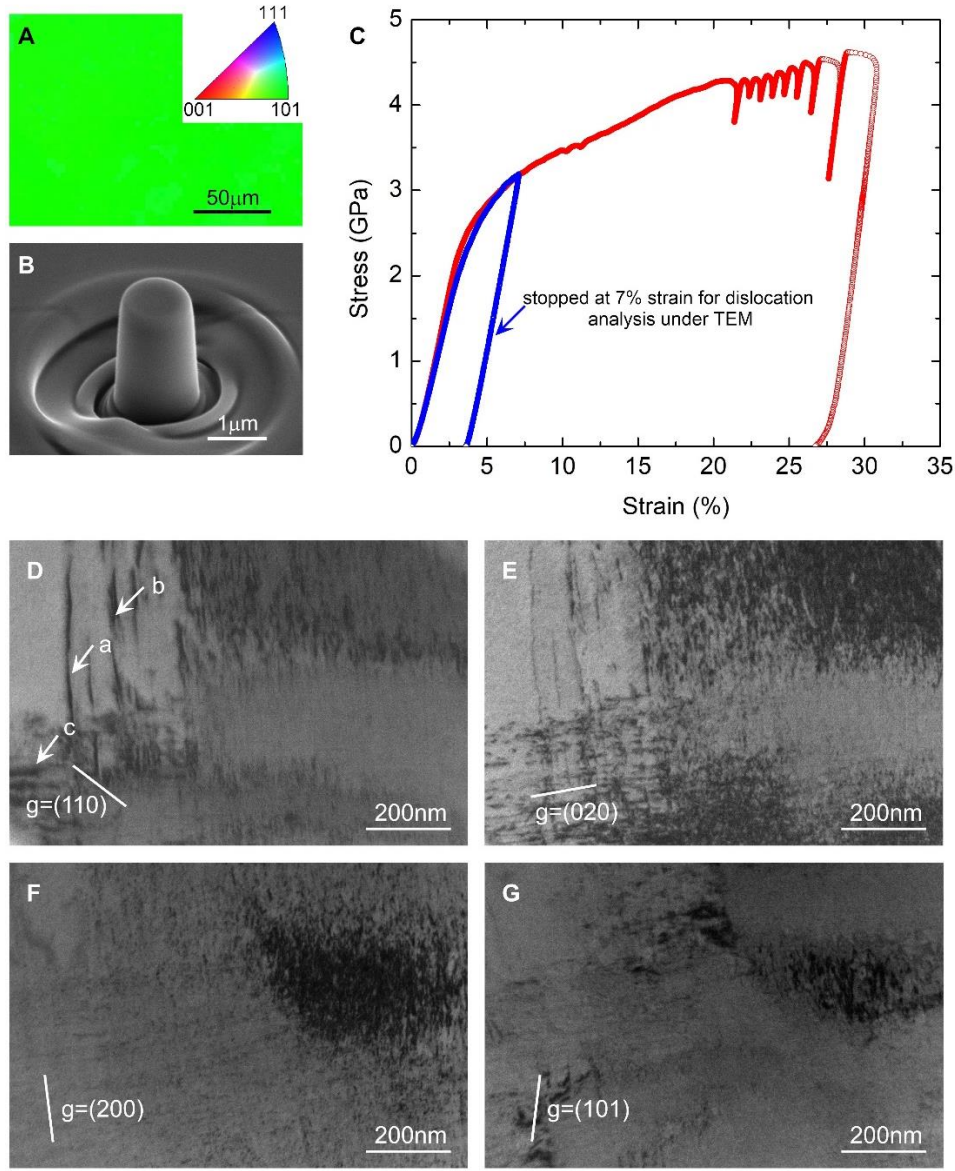

**Fig. S15.**
**The micro compression and dislocation analysis of single crystalline Co25Ni25(HfTiZr)50 along the [110] direction.** (**A**) EBSD image shows the [110] orientation of single crystal. (**B**) A typical pillar SEM image. (**C**) The red curve shows a typical stress-strain curve of the [110] orientated micropillars. The blue curve corresponds to the micropillar deformed to 7% strain for dislocation analysis. (**D-G**) The TEM images under different two-beam conditions. (**D**) $g = (110)$ with the beam direction $Z = [001]$, (**E**) $g = (020)$ with the beam direction $Z = [001]$, (**F**) $g = (200)$ with the beam direction $Z = [001]$ and (**G**) $g = (101)$ with the beam direction $Z = [\bar{1}11]$. The **g•b**=0 out of contrast analyses indicate that the dislocations are of the <001> type. Please see **Supplementary Table S2** for the detailed analyses.

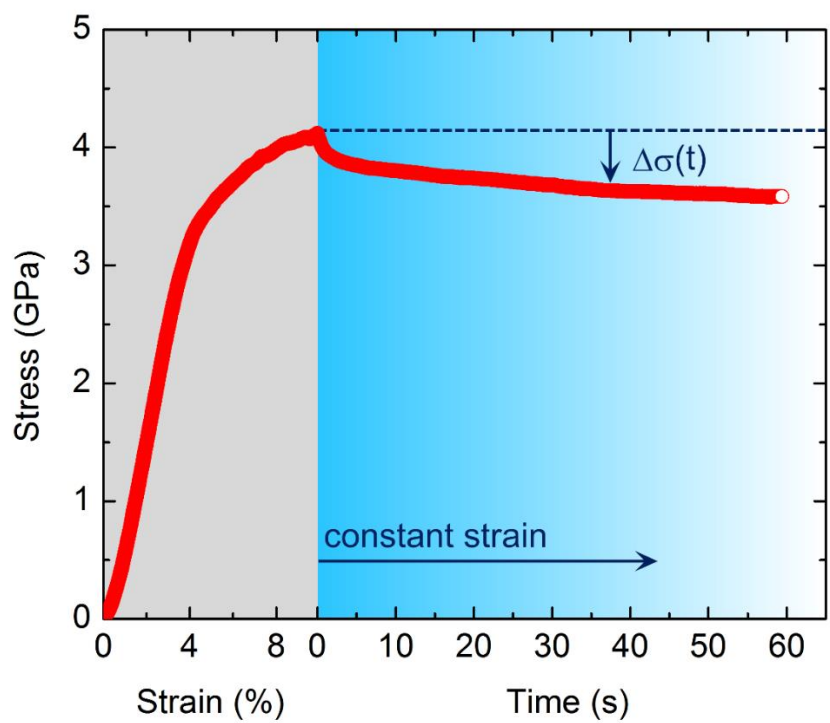

**Fig. S16.**

The typical stress relaxation curve obtained from a single crystalline $Co_{25}Ni_{25}(HfTiZr)_{50}$ micropillar with the top diameter of 1 μm. The activation volume is calculated to be ~3.05b$^3$.

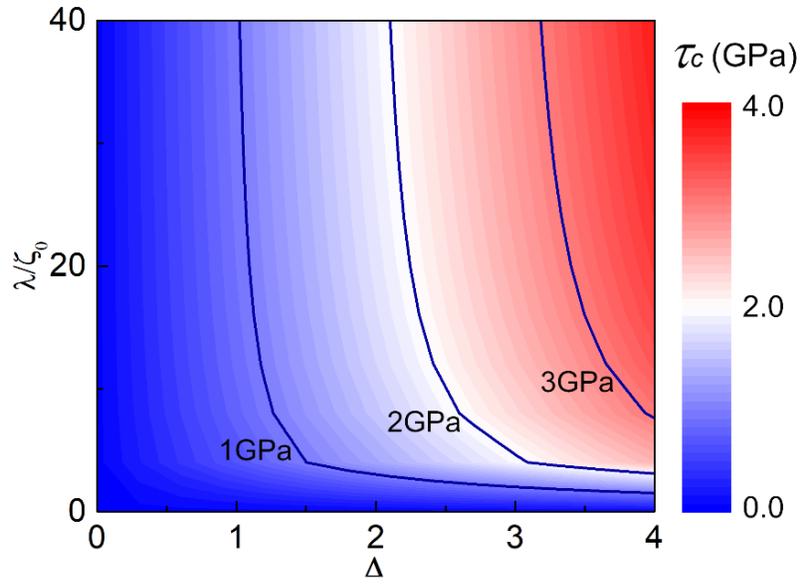

**Fig. S17.**

The contour plot of the critical stress $\tau_c$ as a function of the correlation length $\lambda$ and standard deviation $\Delta$ for $Co_{25}Ni_{25}(HfTiZr)_{50}$.

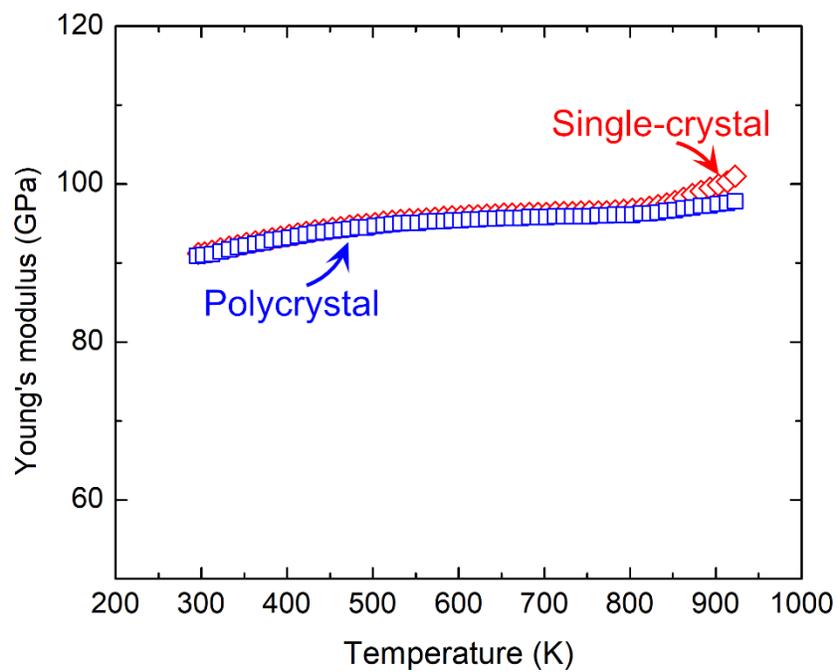

**Fig. S18.**

The Young's modulus versus temperature of the $Co_{25}Ni_{25}(HfTiZr)_{50}$ alloy obtained by the resonance technique.

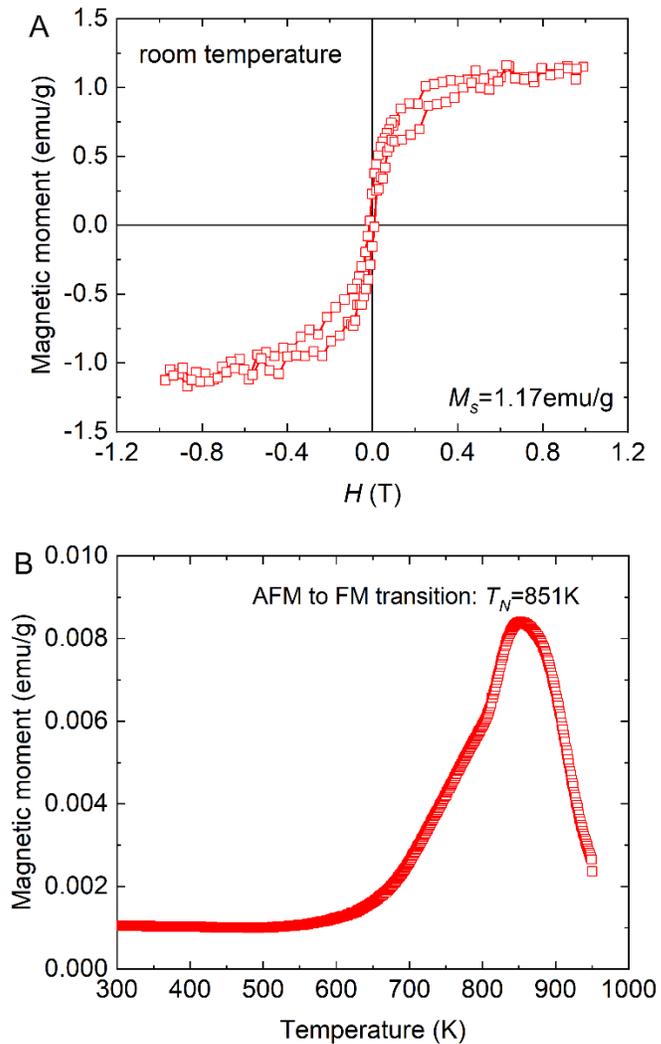

**Fig. S19.**

**The magnetic properties measured for the Co₂₅Ni₂₅(HfTiZr)₅₀ alloy.** (**A**) The magnetization curve of the Co₂₅Ni₂₅(HfTiZr)₅₀ alloy as a function of the applied magnetic field at room temperature. The saturation magnetization $M_s$ of Co₂₅Ni₂₅(HfTiZr)₅₀ is only 1.17 emu/g. (**B**) The temperature dependence of magnetization of Co₂₅Ni₂₅(HfTiZr)₅₀ under the applied magnetic field of 500 Oe. The result shows that there is an antiferromagnetic (AFM) to ferromagnetic (FM) transition at the transition temperature $T_N$=851 K.

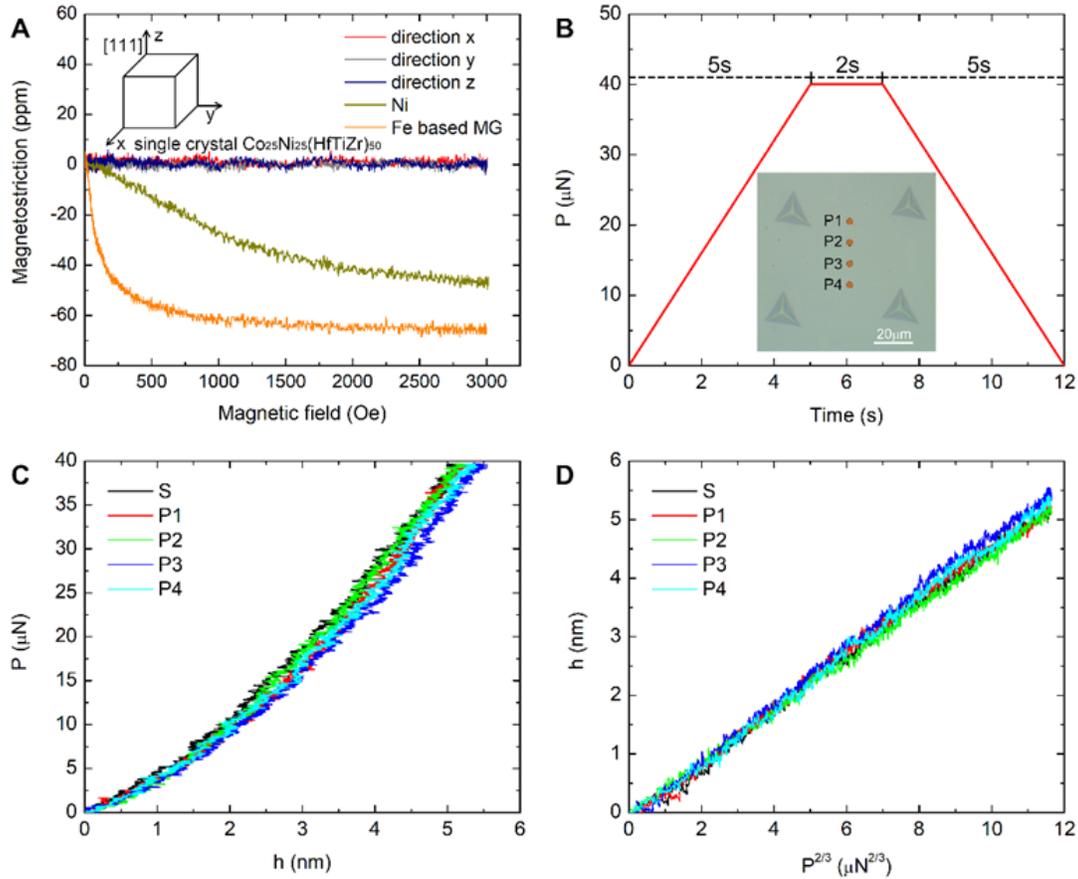

**Fig. S20.**

**The study of magnetostriction behavior of the Co₂₅Ni₂₅(HfTiZr)₅₀ alloy.** (**A**) The measured magnetostriction coefficient along different directions of single crystal $Co_{25}Ni_{25}(HfTiZr)_{50}$ alloy. The magnetostriction coefficient of $Co_{25}Ni_{25}(HfTiZr)_{50}$ alloy is about zero. The ferromagnetic polycrystalline Ni and Fe based metallic glass (MG) are taken for comparison. The measured saturation magnetostriction of Ni is consistent with Ref. (*26*). (**B-D**) The study of magnetostriction obtained using the method of Ref. (*27*). (**B**) The load-time function for the low load spherical indentation with tip radius ≅300nm. Inset is an optical microscopic image of the regular indentation pattern on the $Co_{25}Ni_{25}(HfTiZr)_{50}$ surface (indentation spacing = 60μm), reated with a Berkovich tip and a peak load of 1N. (**C**) The nanoindentation unloading curves. The position S is far from the high load indentation deformation influence zone and the positions of P1, P2, P3 and P4 are labeled in (**B**). (**D**) The $P^{2/3}$ versus $h$ unloading curves. The intercepts of linear fitting represent the displacement induced by magnetostriction $h_{mag}$ which is zero within the (small) measurement error (*27*). This confirms that the magnetostriction effect is VERY weak in $Co_{25}Ni_{25}(HfTiZr)_{50}$.

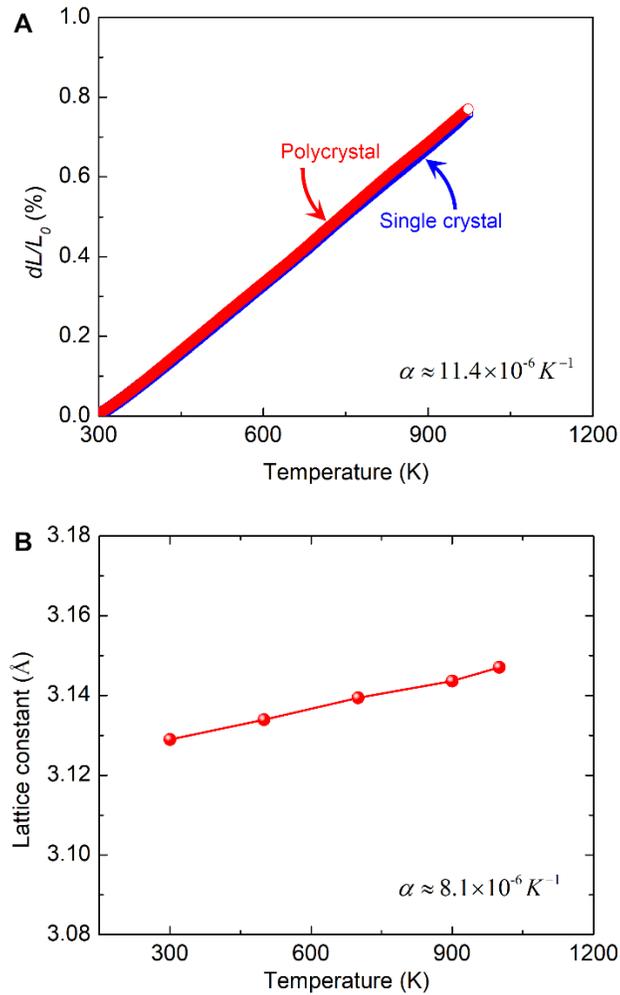

**Fig. S21.**

**The linear thermal expansion coefficient of the Co$_{25}$Ni$_{25}$(HfTiZr)$_{50}$ alloy.** (**A**) The thermal expansion curves obtained from experiments. The average thermal expansion coefficients of the single- and poly-crystalline samples are almost the same and about $11.4 \times 10^{-6}$ K$^{-1}$. (**B**) The variation of the lattice constant with temperature calculated from the ab initio molecular dynamics simulations. The average thermal expansion coefficient is about $8.1 \times 10^{-6}$ K$^{-1}$, which is very close to our experimental measurement.

**Table S1.**

The **g•b**=0 extinction condition analysis of the dislocations marked in **Supplementary Fig. S14**. All the dislocations are of the <001> type. The letters "V" and "I" in the following table represent "visible" and "invisible" respectively.

| Zone axis | g | Dislocations | | | |
|---|---|---|---|---|---|
| | | a | b | c | d |
| [001] | (0$\bar{2}$0) | V | V | I | I |
| [001] | ($\bar{1}\bar{1}$0) | V | V | V | V |
| [001] | (200) | I | I | V | V |
| [$\bar{1}$11] | (0$\bar{1}$1) | V | V | I | I |
| [$\bar{1}$11] | ($\bar{1}\bar{1}$0) | V | V | V | V |
| [$\bar{1}$11] | (10$\bar{1}$) | I | I | V | V |
| **Possible burgers vector** | | [010] or [0$\bar{1}$0] | | [100] or [$\bar{1}$00] | |

**Table S2.**

The **g•b**=0 extinction condition analysis of the dislocations marked in **Supplementary Fig. S15**. All the dislocations are of the <001> type. The letters "V" and "I" in the following table represent "visible" and "invisible" respectively.

| Zone axis | g | Dislocations | | |
|---|---|---|---|---|
| | | a | b | c |
| [001] | (110) | V | V | V |
| [001] | (020) | V | V | V |
| [001] | (200) | I | I | I |
| [$\bar{1}$11] | (10$\bar{1}$) | I | I | I |
| **Possible burgers vector** | | [010] or [0$\bar{1}$0] | | |

**Table S3.**

A summary of the elastic properties and internal friction of shape memory alloys. (AC: as-cast, HT: heat treatment, CR: cold rolled, HR: hot rolled.) Here $^w/_o$ stands for weight percentage while $^a/_o$ for atomic percentage.

| Composition | Processing | E (GPa) | $\sigma_y$ (MPa) | Elastic limit | Internal friction | Ref. |
|---|---|---|---|---|---|---|
| FeMn28Si6Cr5Nb0.53C0.06 $^w/_o$ | AC+HT | 160 | | | 0.01 | (28) |
| FeMn28Si6Cr $^w/_o$ | | 170 | 270 | ~2% | | (29) |
| FeMn17Si5Cr10Ni4VC1 $^w/_o$ | AC+HT | 160 | 415 | 0.26% | | (30) |
| FeMn14.6Si4.22Cr7.98Ni4.14C0.012 $^w/_o$ | AC+HT | | 300 | | | (31) |
| Ni50Ti50 $^a/_o$ | CR | ~56 | 900 | 1.6% | 0.0095 | (32) |
| | CR+HT(300℃) | ~47 | 160 | 0.3% | 0.01 | (32) |
| | CR+HT(500℃) | ~54 | 125 | 0.23% | 0.014 | (32) |
| | CR+HT(600℃) | ~65 | 110 | 0.169% | 0.009 | (32) |
| | CR+HT(700℃) | ~67 | 100 | 0.149% | 0.007 | (32) |
| Ti50Ni25Cu25 $^a/_o$ | HT | | | ~2% | ~0.016 | (33, 34) |
| Ti50Ni50 $^a/_o$ | AC+CR+HT | ~40 | | | 0.065 | (35) |
| Ni49.8Ti42.2Hf8 $^a/_o$ | AC+CR+HT | | | 1.2% | ~0.007 | (36) |
| Ni40Ti50Cu10 $^a/_o$ | AC+HT | 35 | | | ~0.01 | (37) |
| Ni30Ti50Cu20 $^a/_o$ | AC+HT | 34 | | | ~0.011 | (38) |
| CuAl13.5Ni4 $^w/_o$ | AC+HT | ~37.5 | | | 0.008 | (39) |
| CuZn13Al9 $^w/_o$ | AC+HT | | | | 0.02 | (40) |
| Cu72.9Zn20Al5.6Mn1Zr0.5 $^w/_o$ | AC | | | ~0.5% | | (41) |
| Cu73Al17Mn10 $^a/_o$ | AC+CR+HT | | 90-110 | | 0.03-0.043 | (42) |
| (Cu72.8Al17Mn10.2)99.8B0.2 $^a/_o$ | AC+CR+HT | | 184-240 | | 0.025-0.033 | (42) |
| $(Cu_{74}Al_{17.5}Mn_{8.5})_{98}Ni_2$ a/o | AC+CR+HT | | 47 | | 0.092 | (*42*) |
| $(Cu_{74.5}Al_{17.5}Mn_8)_{97.9}Ni_2Si_{0.1}$ | AC+CR+HT | | 225 | | 0.042 | (*42*) |
| CuAlMn | AC+HT | | | ~0.3% | 0.013 | (*43*) |
| CuAlMn+0.4%CuZr doping w/o | AC+HT | | | ~0.5% | 0.022 | (*43*) |
| CuAlMn+0.5%CuZr doping w/o | AC+HT | | | ~0.45% | 0.0225 | (*43*) |
| CuAlMn+0.7%CuZr doping w/o | AC+HT | | | ~0.45% | 0.023 | (*43*) |
| CuAlMn+0.9%CuZr doping w/o | AC+HT | | | ~0.7% | 0.026 | (*43*) |
| CuAlMn+1.0%CuZr doping w/o | AC+HT | | | ~0.5% | 0.02 | (*43*) |
| $In_{51.7}Sn_{48.3}$ w/o | Powder sintering | 28.9 | | | 0.28 | (*44*) |
| $In_{90}Sn_{10}$ w/o | Powder sintering | 27.6 | | | 0.075 | (*44*) |
| $Ti_{50}Pd_{40}Cr_{10}$ w/o | AC | ~28 | | | 0.02 | (*45*) |
| $Ni_{2.15}Mn_{0.81}Fe_{0.04}Ga$ | AC | | | | ~0.015 | (*46*) |

**Table S4.**

A summary of the elastic properties and internal friction of gum metals. (AC: as-cast, HT: heat treatment, CR: cold rolled, HR: hot rolled.)

| Composition | Processing | E (GPa) | $\sigma_y$ (MPa) | Elastic limit | Internal friction | Ref. |
|---|---|---|---|---|---|---|
| Ti12Ta9Nb3V6Zr1.5O $^{a}/_{o}$ | 90% Cold worked | 57.0 | 1102 | 2.5% | | (47) |
| | Before cold worked | 68.4 | 852 | 1.2% | 0.016 | (47, 48) |
| Ti-24Nb-2Zr-7.5Sn $^{w}/_{o}$ | AC+HT | 63.5 | 265 | 0.4% | | (49) |
| Ti-24Nb-4Zr-7.5Sn $^{w}/_{o}$ | AC+HT | 51.7 | 221 | 0.4% | | (49) |
| Ti-24Nb-4Zr-11.5Sn $^{w}/_{o}$ | AC+HT | 79.4 | 601 | 0.8% | | (49) |
| Ti-23Nb-0.7Ta-2Zr-1.2O $^{w}/_{o}$ | HT+CR | 60 | 808 | 1.4% | | (50) |
| Ti-22.5Nb-0.7Ta-2Zr $^{a}/_{o}$ | HR | 54.2 | 285 | 0.5% | | (51) |
| Ti-22.5Nb-0.7Ta-2Zr-0.5O $^{a}/_{o}$ | | 57.1 | 708 | 1.2% | | (51) |
| Ti-22.5Nb-0.7Ta-2Zr-1.0O $^{a}/_{o}$ | | 63.3 | 657 | 1.0% | | (51) |
| Ti-22.5Nb-0.7Ta-2Zr-1.5O $^{a}/_{o}$ | | 59.0 | 782 | 1.3% | | (51) |
| Ti-22.5Nb-0.7Ta-2Zr-2.0O $^{a}/_{o}$ | | 91.0 | 814 | 0.9% | | (51) |
| Ti-26.53Nb-10.14Ta-1.63Zr-0.28Fe-0.13O-0.01C-0.007N-0.001H $^{w}/_{o}$ | HR | 75 | 500 | 0.93% | | (52) |
| | HR+CR 30% | 73 | 662 | 1.39% | | (52) |
| | HR+CR 60% | 68 | 787 | 1.78% | | (52) |
| | HR+CR 80% | 64 | 821 | 1.68% | | (52) |

**Table S5.**

The geometrical data for the micropillars. SE indicates pillars used to study size effect (Fig. 3B)

and HT indicates pillars used to study the temperature dependence of the yield strength (Fig. 3D).

| Pillar No. | Diameter(μm) | Height(μm) | Aspect ratio | Taper angle |
|------------|--------------|------------|--------------|-------------|
| SE1 | 5.19 | 12.78 | 2.40 | 2.40 |
| SE2 | 5.17 | 11.90 | 2.06 | 2.06 |
| SE3 | 5.11 | 11.44 | 2.35 | 2.35 |
| SE4 | 4.63 | 9.61 | 2.08 | 2.34 |
| SE5 | 4.57 | 10.10 | 2.21 | 2.25 |
| SE6 | 4.52 | 10.12 | 2.24 | 2.29 |
| SE7 | 4.40 | 9.94 | 2.26 | 2.19 |
| SE8 | 4.16 | 10.02 | 2.03 | 2.03 |
| SE9 | 4.12 | 11.24 | 2.29 | 2.29 |
| SE10 | 3.91 | 6.01 | 1.54 | 3.52 |
| SE11 | 3.89 | 5.99 | 1.54 | 3.88 |
| SE12 | 3.79 | 6.26 | 1.65 | 3.64 |
| SE13 | 3.20 | 8.01 | 2.22 | 2.22 |
| SE14 | 3.13 | 7.81 | 2.20 | 2.20 |
| SE15 | 3.13 | 10.51 | 2.30 | 2.30 |
| SE16 | 2.98 | 6.43 | 2.16 | 1.98 |
| SE17 | 2.96 | 6.51 | 2.20 | 1.84 |
| SE18 | 2.95 | 6.57 | 2.23 | 2.05 |
| SE19 | 2.94 | 6.15 | 2.09 | 2.18 |
| SE20 | 2.94 | 6.49 | 2.21 | 1.61 |
| SE21 | 2.94 | 6.29 | 2.14 | 1.99 |
| SE22 | 2.90 | 6.11 | 2.10 | 2.31 |
| SE23 | 2.28 | 6.46 | 2.67 | 2.67 |
| SE24 | 2.22 | 7.29 | 2.56 | 2.56 |
| SE25 | 2.15 | 5.93 | 2.90 | 2.90 |
| SE26 | 1.25 | 2.34 | 2.42 | 2.42 |
| SE27 | 1.21 | 2.19 | 2.71 | 2.71 |
| SE28 | 1.19 | 2.55 | 2.98 | 2.98 |
| SE29 | 1.03 | 2.19 | 2.14 | 3.62 |
| SE30 | 0.99 | 2.14 | 2.17 | 3.26 |
| SE31 | 0.96 | 2.24 | 2.34 | 3.30 |
| SE32 | 0.54 | 1.21 | 3.75 | 3.75 |
| SE33 | 0.52 | 1.25 | 3.80 | 3.80 |
| SE34 | 0.52 | 1.26 | 3.23 | 3.23 |
| SE35 | 0.35 | 0.78 | 2.22 | 3.72 |
| SE36 | 0.34 | 0.87 | 2.55 | 3.13 |
| SE37 | 0.31 | 0.78 | 2.51 | 3.59 |
| SE38 | 0.30 | 0.92 | 3.06 | 3.60 |
| SE39 | 0.28 | 0.58 | 2.08 | 3.25 |
| HT1 | 1.18 | 2.52 | 2.14 | 2.05 |
| HT2 | 1.14 | 2.27 | 1.99 | 2.19 |
| HT3 | 1.24 | 2.44 | 1.96 | 2.77 |
| HT4 | 1.23 | 2.40 | 1.94 | 2.60 |
| HT5 | 1.24 | 2.47 | 2.00 | 2.52 |
| HT6 | 1.21 | 2.57 | 2.12 | 2.32 |
| HT7 | 1.20 | 2.71 | 2.25 | 2.33 |
| HT8 | 1.22 | 2.26 | 1.86 | 2.06 |
| HT9 | 1.21 | 2.61 | 2.15 | 2.22 |
| HT10 | 1.23 | 2.61 | 2.13 | 2.21 |
| HT11 | 1.21 | 2.82 | 2.33 | 1.96 |
| HT12 | 1.21 | 2.55 | 2.10 | 2.39 |
| HT13 | 1.21 | 2.97 | 2.45 | 2.00 |
| HT14 | 1.22 | 2.93 | 2.41 | 2.28 |
| HT15 | 1.25 | 2.59 | 2.07 | 2.57 |
| HT16 | 1.24 | 2.46 | 1.98 | 3.08 |
| HT17 | 1.23 | 2.40 | 1.96 | 2.81 |
| HT18 | 1.22 | 2.55 | 2.08 | 2.92 |
| HT19 | 1.20 | 2.62 | 2.19 | 2.87 |
| HT20 | 1.26 | 2.27 | 1.81 | 3.38 |
| HT21 | 1.21 | 2.43 | 2.00 | 2.71 |

**Movie S1.**

The visualization of ultra-elasticity in the $Co_{25}Ni_{25}(HfTiZr)_{50}$ alloy through the steel ball dropping tests. Several other alloys are also included in the video for comparison.